\documentclass[12pt]{mn2e}
\usepackage{graphicx}
\usepackage{epsf}
\usepackage{lscape}
\usepackage{longtable}
\usepackage{rotating}
\usepackage{color}
\usepackage{comment}

\newcommand{\gtrsim}{\ga}

\newcommand{\apj}{ApJ}
\newcommand{\apjl}{ApJ}

\newcommand{\mnras}{MNRAS}

\newcommand{\pasp}{PASP}

\newcommand{\nat}{Nat}

\begin{document}
\topmargin -0.5in 

\title[Spectroscopy of $z>7$ candidate galaxies]{No Evidence for Lyman-$\alpha$ Emission in Spectroscopy of $z>7$ Candidate Galaxies\thanks{Based on observations collected at the European Organisation for Astronomical Research in the Southern Hemisphere, Chile, as part of programme 086.A-0968(B).}}

\author[Joseph\ Caruana, et al.\ ]  
{Joseph Caruana$^{1}$\thanks{E-mail: joseph.caruana@astro.ox.ac.uk}, Andrew J. Bunker$^{1}$, Stephen M. Wilkins$^{1}$, Elizabeth R. Stanway$^{2}$, 
\newauthor
Mark Lacy$^{3}$,
Matt J.\ Jarvis$^{4}$,
Silvio Lorenzoni$^{1}$ \& Samantha Hickey$^{4}$\\
$^1$\,University of Oxford, Department of Physics, Denys Wilkinson Building, Keble Road, OX1 3RH, U.K. \\
$^{2}$\,Department of Physics, University of Warwick, Gibbet Hill Road, Coventry, CV4 7AL, U.K.\\
$^{3}$\,NRAO, 520 Edgemont Road, Charlottesville, VA 22903, USA\\
$^{4}$\,Centre for Astrophysics, Science \& Technology Research Institute, University of Hertfordshire, Hatfield, Herts AL10\,9AB, U.K
.}
\maketitle

\begin{abstract}
We present Gemini/GNIRS spectroscopic observations of 4 $z-$band ($z\approx 7$) dropout galaxies and VLT/XSHOOTER observations of one $z-$band dropout and 3 $Y-$band ($z\approx 8-9$) dropout galaxies in the {\em Hubble} Ultra Deep Field, which were selected with Wide Field Camera 3 imaging on the {\em Hubble Space Telescope}. We find no evidence of Lyman-$\alpha$ emission with a typical $5\sigma$ sensitivity of $5\times10^{-18}$erg cm$^{-2}$s$^{-1}$, and we use the upper limits on Lyman-$\alpha$ flux and the broad-band magnitudes to constrain the rest-frame equivalent widths for this line emission. Accounting for incomplete spectral coverage, we survey 3.0 $z$-band dropouts and 2.9 $Y$-band dropouts to a Lyman-$\alpha$ rest-frame equivalent width limit $>120$\,\AA\ (for an unresolved emission line); for an equivalent width limit of $50$\,\AA\ the effective numbers of drop-outs surveyed fall to 1.2 $z$-band drop-outs and 1.5 $Y$-band drop-outs.  A simple model where the fraction of high rest-frame equivalent width emitters follows the trend seen at $z=3-6.5$ is inconsistent with our non-detections at $z=7-9$   at the $\approx 1\sigma$ level for spectrally unresolved lines, which may indicate that a significant neutral HI fraction in the intergalactic medium suppresses the Lyman-$\alpha$ line in $z$-drop and $Y$-drop galaxies at $z>7$.
\end{abstract} 


\section{Introduction}

The Lyman break technique has proven to be an efficient tool in the selection of high-redshift candidate galaxies - and rest-frame UV-selected galaxies are now being regularly identified at redshifts $z\gtrsim6$ (e.g. Bunker et al.\ 2004; Bouwens et al.\ 2006; McLure et al.\ 2010).  With the advent of the Wide Field Camera 3 (WFC3) on the {\em Hubble} Space Telescope ({\em HST}), the technique has been pushed to even higher redshifts and the number of candidates beyond $z=6.5$ has increased to about a hundred (e.g. Bouwens et al.\ 2011, McLure et al.\ 2011, Lorenzoni et al.\ 2011, Wilkins et al.\ 2011, Finkelstein et al.\ 2010, Wilkins et al.\ 2010, Bunker et al.\ 2010, Oesch et al.\ 2010a).  Spectroscopic confirmation of these high-$z$ candidate galaxies remains a very important task.  However, the sample of convincing spectroscopically-confirmed sources remains small (eg. Ono et al.\ 2012, Pentericci et al.\ 2011, Schenker et al.\ 2012, Vanzella et al.\ 2011).  Not knowing the precise redshift of most Lyman break candidates by means of spectroscopic confirmation reduces our confidence in inferred properties about the universe at this epoch, both due to redshift uncertainties on each source and because of the unquantifiable risk of contamination in these samples (e.g. see Hayes et al.\ 2012).  For objects at such high redshifts, the only currently-feasible spectroscopic redshift diagnostic is the Lyman-$\alpha$ emission line, which arises from photoionization of HII regions by star formation.  The line itself is resonant and sensitive to the ionization state of the intergalactic medium (IGM) and thus, its visibility at the end of cosmic reionization is expected to be reduced compared to that at lower redshifts due to the damping effect of an increasingly neutral IGM (Gunn \& Peterson 1965; Becker et al.\ 2001).  The emergence or non-emergence of Lyman-$\alpha$ may be indicative of the size of the HII regions surrounding these galaxies; in order for Lyman-$\alpha$ emission to emerge from a galaxy, the ionized region surrounding the galaxy needs to be sufficiently large to allow the wavelength of the Lyman-$\alpha$ photon to redshift to longer wavelengths, and hence become non-resonant when encountering the neutral IGM, thus managing to escape.

In order to address the question of the emergence of Lyman-$\alpha$ at these high redshifts, we undertook Gemini/GNIRS and VLT/XSHOOTER spectroscopy of a sample of 7 Lyman-break selected candidate galaxies at $z\gtrsim7$, identified as $z$-band and $Y$-band dropouts with {\em HST}/WFC3 imaging (Bunker et al.\ 2010, Lorenzoni et al.\ 2011, Wilkins et al.\ 2011) centered on and around the Hubble Ultra Deep Field (HUDF, Beckwith et al.\ 2006).  We have already described the analysis of one of these objects (HUDF.YD3) in a separate paper (Bunker et al.\ 2012), following the reported detection of Lyman-$\alpha$ at $z=8.55$ by Lehnert et al.\ (2010), so we do not discuss it again here.

The structure of this paper is as follows. We describe our spectroscopic observations and data reduction in Section \ref{sec:obs}, and present the results of the spectroscopy in Section \ref{sec:results}.  We discuss our analysis in Section \ref{sec:analysis} and our conclusions are summarized in Section \ref{sec:conc}.  We adopt a $\Lambda$CDM cosmology throughout, with $\Omega_M=0.3$, $\Omega_{\Lambda}=0.7$ and $H_0=70\,{\rm km\,s^{-1}\,Mpc^{-1}}$. All magnitudes are in the AB system (Oke \& Gunn 1983).

\section{OBSERVATIONS AND DATA REDUCTION}
\label{sec:obs}

\subsection{Observations with Gemini/GNIRS}

Four of these objects had initially been selected as candidate $z>7$ sources with {\em HST}/NICMOS imaging as they were undetected in the ACS images of the HUDF (Bunker et al.\ 2004) and re-confirmed with {\em HST}/WFC3 (Bunker et al.\ 2010, Oesch et al.\ 2010a).  All four of these objects have subsequently appeared in the independent selection of several different groups (see Wilkins et al.\ 2011 for a comparison).  We also obtained spectroscopy on 5 additional candidate galaxies at $z>7$ that were selected with {\em HST}/NICMOS but were later rejected as secure candidates by the deeper {\em HST}/WFC3 imaging.  For this reason, we do not include these 5 objects in our current analysis.

We observed the four $z$-band drop-out high-redshift galaxy candidates HUDF.zD1, HUDF.zD2, HUDF.zD3 and HUDF.zD4 from Bunker et al.\ (2010), hereafter called zD1, zD2, zD3 and zD4.  We used the GNIRS spectrograph on the Gemini South telescope as part of programmes GS-2004B-Q-19 and GS-2005B-Q-18 (PI: A. Bunker, see also Stanway et al.\ 2004a).  GNIRS is a spectrometer that can perform both long-slit, single-order spectroscopy in the $1.0-5.4\mu$m range and cross-dispersed spectroscopy in the $0.9-2.5\mu$m range.  We used the latter cross-dispersed mode with the ``high-dispersion'' 110.5 l/mm grating and short blue camera, with a 7$''$-long slit.  In this mode, we do not get continuous spectral coverage since not all of the spectral orders fall on the array.

Three of our objects (zD2, zD3 and zD4) were observed using a grating angle of 21 degrees and the XD\_G0525 filter, which allowed us to include the regions 0.86-0.94$\mu$m and 1.005-1.095$\mu$m in our range of covered wavelengths, corresponding to redshifts $z\sim7$ for Lyman-$\alpha$ (as expected for $z$-band dropouts).  The observations of zD1 were carried out using two observing settings.  For one setting, we used the same grating angle of 21 degrees but used the XD\_G0507 filter instead, whereas for the second setting we changed the grating angle to 23.2 degrees and used the XD\_G0525 filter.  This allowed us to fill the missing wavelength range between 0.94$\mu$m and 1.005$\mu$m, as well as covering the region 0.83$\mu$m-0.90$\mu$m, thus having continuous spectroscopy between 0.83$\mu$m and 1.095$\mu$m ($z=5.8-8.0$ for Lyman-$\alpha$).

The coordinates for these four objects are taken from Bunker et al.\ (2010).  
In acquiring the object, a nearby star (with accurately determined astrometry from {\em HST}) was first centered in the slit and then a blind offset (less than $1'$) was executed to place the faint $z$-dropout galaxy in the slit. The Gemini telescope can execute an offset of this size to an accuracy of better than $0\farcs2$\footnote{See \tt http://www.gemini.edu/sciops/telescopes-and-sites/\\acquisition-hardware-and-techniques?\%20q=node/10769} when the same guide star is used throughout (as was the case for our observations).  A slit width of 0\farcs675 was used for all observations and the data were acquired in a three-point dither pattern (ABC) at positions +1\farcs8, 0, -1\farcs8 along the slit long axis. We simulated the effect of offset errors when calculating the slit losses, and concluded this had at most a 10 per cent effect on the flux captured by the aperture. Each frame was a 900s exposure.  In total, 9 frames were combined for each setting of zD1, and also for zD3 and zD4, whilst 7 frames were combined for zD2.
The observations were conducted between 2004 and 2006 (see Table \ref{object_table}).    The seeing measured from the point spread function (PSF) of a star in the acquisition spectrum was about $0\farcs5'$ Full Width at Half Maximum (FWHM).  The seeing was monitored by the observers in real time via the Peripheral Wavefront Sensor star, and if the seeing degraded then the queue-mode observations of this programme were terminated and observations switched to a programme with a less stringent seeing requirement. In some instances during the observation of our target for extended periods, a reacquisition of the blind offset star was performed. We confirmed that the blind offset star appeared in the expected position (i.e.\ the telescope pointing had not wandered off), and we also checked that the seeing was again consistent (about $0\farcs5$ on average). Hence we are confident of the seeing used to estimate the slit losses. We carried out flux calibration using observations of a spectrophotometric A0V telluric standard star which was observed on the same dates as the objects.  From unblended spectral lines in the sky spectra we measured a spectral resolving power of $R=\lambda/\Delta\lambda_{\rm FWHM}=3700$.  

We reduced the GNIRS spectra with {\tt IRAF} using our own customized reduction pipeline, which is partly based on an existing Gemini pipeline.  We first used the {\tt IRAF} GNIRS task {\tt nvnoise} to remove a vertical striping pattern that arises from offset bias levels.  We then used {\tt nsprepare} to create variance and data quality planes for our frames.  We performed a first-pass of cosmic ray rejection using the {\tt IRAF} tasks {\tt imcombine} and {\tt ccdclip} and input CCD noise and gain parameters.  We then used the resulting rejection frames to mask those pixels affected by cosmic ray strikes.  We used the same task a second time to apply a further pass of cosmic ray rejection, also using the data quality array to mask bad pixels.  Flat-field generation was achieved by using the tasks {\tt nsreduce} and {\tt nsflat} on three sets of flats - each appropriate for different orders:  the first set being IR lamp flats, the second set being short-exposure Quartz Halogen (QH) lamp flats and the third set being long-exposure QH lamp flats.  Subsections of the data corresponding to each spectral order were then trimmed out from the original 2D spectrum and these were sky-subtracted to first order by subtracting the average of the other dither positions and flat-fielded using {\tt nsreduce}.  We rectified the spectra (spatially and spectrally) using two-dimensional arc-spectra through a pinhole file, correcting for geometric distortion of the optics in the detector, and removed any remaining skyline residuals using the {\tt background} task.  We then added on a model of the skylines to every frame prior to combining them with {\tt imcombine} to enable the rejection of any remaining cosmic ray strikes using a Poisson noise model.  Finally, the sky model was subtracted from the resulting frame.  We obtained flux calibration using the above mentioned telluric standard star by means of standard {\tt IRAF} techniques.  The star was extracted using the {\tt apall} task.  We used {\tt standard}  to integrate the observations of the standard stars over calibration bandpasses, obtain a flux using a blackbody flux distribution model and apply an extinction correction.  We then used the task {\tt sensfunc} to determine the system sensitivity and finally we used the {\tt calibrate} task to flux calibrate the data.

\subsection{Observations with VLT/XSHOOTER}

We observed the two $Y$-band dropouts HUDF.YD3\footnote{This corresponds to UDFy-38135539 in Bouwens et al.\ (2010), 1721y in McLure et al.\ (2010), z8-B115 in Yan et al.\ (2010), object 125 in Finkelstein et al.\ (2010)} (Lorenzoni et al.\ 2011) and ERS.YD2 (Lorenzoni et al.\ 2011.) and the $z$-band dropout P34.z.4809\footnote{This corresponds to the $Y$-drop UDF092y-03781204 in Bouwens et al.\ 2011.} (Wilkins et al.\ 2011) using the XSHOOTER spectrograph (D'Odorico et al.\ 2006) on the ESO VLT-UT2 (Kueyen) as part of programme 086.A-0968(B) (PI: A. Bunker). XSHOOTER is an echelle spectrograph, with UV, visible and near-infrared channels obtaining near-continuous spectroscopy from $0.3\mu$m to $2.48\mu$m with a 1$\farcs2$-wide and 11$''$-long slit.  HUDF.YD3 has already been discussed in detail in Bunker et al.\ (2012), so here we shall focus our attention on our spectroscopy of the other targets.  We dithered the observations in an ABBA sequence.  For HUDF.YD3 and ERS.YD2, we dithered at positions $+3''$ and $-3''$ from the central coordinates along the slit long axis (i.e. a `chop' size of $6''$).  In the case of P34.z.4809 we dithered $+2\farcs5$ and $-2\farcs5$ from the central coordinates along the slit long axis (i.e. a `chop' size of $5''$).

For ERS.YD2 we set the central coordinates to those from Lorenzoni et al.\ (2011) - see Table \ref{object_table2}.  We set the position angle (P.A.) to be $+31^\circ$ (measured East of North).  We first peaked up on a nearby star $68\farcs51$ East and $25\farcs7$ South of the desired central pointing, then did a blind offset.
For P34.z.4809 we set the central coordinates to be RA=03:33:03.765 Dec.=-27:51:20.11 (J2000) so that we could target both the position of P34.z.4809 in Wilkins et al.\ (2011) - see Table \ref{object_table2}, and UDF092y-03751196d from Bouwens et al.\ (2010) which is 0\farcs8 away.  The P.A. was set to -22.5$^\circ$ to intercept both objects.  We first peaked up on a nearby star $17\farcs98$ East and $10''$ South of the desired central pointing, and then again did a blind offset. For blind offsets of this size, ESO guarantee an accuracy better than $0\farcs1$ where the guide star remains the same. We simulated the effect of this positional uncertainty on the fraction of the light falling down the slit, and concluded this was at most a $5-10$ per cent effect.

The ERS.YD2 observations were conducted in 6 observing blocks, each including 49mins of on-source integration and consisting of a single ABBA sequence with three exposures of the near-IR arm of duration 245 s at each A or B position. The  observations were taken on the nights of UT 2010 December 07 and 2011 January 02, 04, 05, 11 \& 23.  The vast majority of the frames were taken in good seeing conditions of $0\farcs56-0\farcs76$ FWHM. The P34.z.4809 observations were conducted in 5 observing blocks, 49mins of which consisted of on-source integration.  These were taken on the nights of UT 2010 October 16, 17, 19 \& 28 with two observing blocks taken on the night of UT 2010 October 17 and single observing blocks on the other nights.  Seeing conditions were similar to those during the observations of ERS.YD2.  All observations were taken at low airmass, with an average airmass of 1.16, and 83 per cent of the observations were taken at airmasses below 1.3. At such low airmass, the effect of differential atmospheric dispersion is negligible between the red end of the optical channel and the near-infrared $1-3\,\mu$m (where we expect Lyman-$\alpha$), and we confirmed that the alignment star was centred in the slit in both the optical and near-infrared spectra. Three piezo controlled mirrors, located in front of each arm, guarantee that the optical path is maintained aligned against instrument flexure and correct for differential atmospheric refraction between the telescope guiding wavelength and each arm central wavelength.

The resolving power attained for our IR-channel observations of ERS.YD2 was $R=5000$.  In the case of P34.z4809 we also considered data acquired with the optical arm as well as that from the near-infrared since this object is a $z$-drop, so the expected redshift for Lyman-$\alpha$ range extends down to just below 1$\mu$m.  These optical data were obtained simultaneously with the IR data and were acquired in 5 observing blocks, each including 49 minutes of on-source integration.  The resolving power for the optical channel was $R=6700$. There is a small region of overlap between the two arms ($0.994-1.013\,\mu$m) and in this range we combined the 2D spectra using inverse-variance weighting as a function of wavelength, which accounted for the strong variation of throughputs/sensitivities with wavelength in the overlap region and also the different readout noise characteristics of the two detectors.

We used the ESO pipeline (Modigliani et al.\ 2010) to reduce our data.  This pipeline applied spatial and spectral rectification to the spectra (which exhibited significant spatial curvature as well as a non-linear wavelength scale) by using the two-dimensional arc spectra through a pinhole mask.  For the IR channel, the pipeline mapped the data to an output spectral scale of 1\,\AA\,pix$^{-1}$ and a spatial scale of  0\farcs21 (from original scales of about 0.5\,\AA\,pix$^{-1}$ and 0\farcs24 respectively).  For the $z$-drop P.34.z.4809 we might expect that Lyman-$\alpha$ falls at the red end of the optical spectrum (around 0.9\,$\mu$m), hence for this object we also inspected the optical channel, where the data were mapped to an output spectral scale of 0.4\,\AA\,pix$^{-1}$ and a spatial scale of 0\farcs16.  In both channels the data were flat-fielded and cosmic rays were identified and masked using the algorithm of van Dokkum (2001).  The two dither positions were subtracted to remove the sky to first order, and the different echelle orders were combined together into a continuous spectrum (taking into account the variation in throughput with wavelength in different overlapping echelle orders) before spatially registering and combining the data taken at the two dither positions, and removing any residual sky background. Flux calibration was achieved through observations of standard stars LTT3218, GD-71 and Feige 110 taken on the same nights as the science data. We chose the subset of the standard star observations that were taken in similar seeing conditions and at similar airmass to our science data.

\begin{table*}
\caption{$z$-band dropouts targetted with Gemini/GNIRS.  The RA \& Dec positions are from Bunker et al.\ (2010). The $Y$- and $J$-band magnitudes quoted for zD1, zD2 and zD3 are from Wilkins et al.\ (2010) using the WFC3 $Y_{098m}$/$Y_{105w}$ and $F125W$ filters respectively whereas the $Y$- and $J$-band magnitudes quoted for zD4 are from Bunker et al.\ (2010).  $M_{1600}$ is the absolute rest-frame UV magnitude around 1600\AA\ for the most probable redshift.}
\begin{tabular}{| c | c | c | c | c | c | c | c |}
\hline
Object & R.A. (J2000) & Dec (J2000) & $Y_{\rm AB}$-mag & $J_{\rm AB}$-mag & $M_{1600}$ & Exp Time (hrs) & Date \\ 
\hline
\hline
zD1 & 03:32:42.56 & -27:46:56.6 & $26.71\pm 0.03$ & $26.44 \pm 0.03$ &  -20.53 & 2.25 & 27 Nov 2004\\
& & & & & & 2.25 & 20 Dec 2005\\
\hline
zD2 & 03:32:38.81 & -27:47:07.2 & $27.48 \pm 0.06$ & $26.90 \pm 0.04$ & -20.18 & 1.75 & 08 Dec 2005\\
\hline
zD3 & 03:32:42.57 & -27:47:31.5 & $27.50 \pm 0.07$ & $27.10 \pm 0.05$ & -19.64 & 2.25 & 17 Dec 2005\\
\hline
zD4 & 03:32:39.55 & -27:47:17.5 & $27.84 \pm 0.09$ & $27.34 \pm 0.05$ & -19.79 & 2.25 & 18 Dec 2005\\
 & & & & & & 2.25 & 30 Jan 2006\\
\hline
\end{tabular} 
\label{object_table} 
\end{table*}

\begin{table*}
\caption{The 3 $Y$-band dropouts and 1 $z$-band dropout targeted with VLT/XSHOOTER.  The RA \& Dec positions are from Lorenzoni et al.\ (2011) for HUDF.YD3 and ERS.YD2, from Wilkins et al.\ (2011) for P34.z.4809 and from Bouwens et al.\ (2011) for UDF092y-03751196d.  The $J$-band magnitudes for each object are quoted from the same respective papers using the WFC3 $F125W$ filter.  $M_{1600}$ is the absolute rest-frame UV magnitude around 1600\AA\ for the most probable redshift.}
\begin{tabular}{| c | c | c | c | c | c | c |}
\hline
Object & R.A. (J2000) & Dec (J2000) & Mag ($J$) & $M_{1600}$ & Exp Time (hrs) & Date \\ 
\hline

\hline
HUDF.YD3 & 03:32:38.135 & -27:45:54.03 & $28.18 \pm 0.13$ &  -19.12 & 0.82 & 27 Dec 2010\\
& & & & & 1.63 & 29 Dec 2010\\
& & & & & 1.63 & 30 Dec 2010\\
& & & & & 0.82 & 31 Dec 2010\\

\hline
ERS.YD2 & 03:32:02.986 & -27:43:51.95 & $26.98 \pm 0.15$ & -20.28 & 0.82 & 07 Dec 2010\\
& & & & & 0.82 & 02 Jan 2011\\
& & & & & 0.82 & 04 Jan 2011\\
& & & & & 0.82 & 05 Jan 2011\\
& & & & & 0.82 & 11 Jan 2011\\
& & & & & 0.82 & 23 Jan 2011\\

\hline
UDF092y-03751196d & 03:33:03.750 & -27:51:20.40 & $26.30 \pm 0.00$ & -20.92 & 0.82 & 16 Oct 2010\\
& & & & & 1.63 & 17 Oct 2010\\
& & & & & 0.82 & 19 Oct 2010\\
& & & & & 0.82 & 28 Oct 2010\\

\hline
P34.z.4809 & 03:33:03.781 & -27:51:20.48 & $26.39 \pm 0.03$ &  -20.65 & 0.82 & 16 Oct 2010\\
& & & & & 1.63 & 17 Oct 2010\\
& & & & & 0.82 & 19 Oct 2010\\
& & & & & 0.82 & 28 Oct 2010\\
\hline
\end{tabular} 
\label{object_table2} 
\end{table*}

\section{RESULTS}
\label{sec:results}
We inspected all the 2D spectra, focusing in particular on the expected location for the target and the wavelength ranges between 0.8 and 1.2 microns where Lyman-$\alpha$ might be expected.  We did this through visual inspection, including examining frames which had been smoothed by means of convolution with a Gaussian with similar FWHM to the spatial seeing and spectral resolution to bring up any faint feature. 
We developed a noise model, based on the Poisson counts of the sky background and dark current, the sensitivity of the detector (as a function of wavelength and position on the array) and the readout noise of the array. Dividing our reduced 2D spectrum by the noise model provided a map of signal-to-noise ratio.
We ran SExtractor (Bertin and Arnouts 1996) on the 2D spectrum after division by our noise model  to identify possible emission lines which might have been missed through visual inspection.  As a check on contamination by spurious sources and noise spikes, we also examined the negative image to confirm that there were no significant detections.

We do not detect any significant line emission in any of the objects in our sample.  To test the recoverability of possible line emission in our spectroscopy, we added fake sources of various intensity in random locations in our 2D spectra and checked if we would have detected these.  We initially took an emission line with an elliptical Gaussian profile, with a FWHM of 200 km\,s$^{-1}$ in the spectral direction, and compact spatially (to match the typical small sizes of the Lyman break population at $z>6$). We then convolved this with another Gaussian to reflect the instrumental spectral resolution and the ground based seeing (i.e.\ FWHM of about $0\farcs5$ and 2.56\,\AA\ for the near-infrared spectroscopy, and 1.4\,\AA\ for the XSHOOTER optical arm). We note that the effect of the strong Lyman-$\alpha$ forest absorption at these redshifts would be expected to absorb the entire blue wing of this line emission, so the actual line width and flux before absorption might be twice as large. To reflect this, we also experimented with introducing a truncated Gaussian before convolution with the instrumental resolution, where the initial FWHM was 400\,km\,s$^{-1}$ and we set the blue half of the profile to be zero (i.e.\ the FWHM is now 200\,km\,s$^{-1}$). This profile produced similar recoverability statistics to the complete Gaussian simulations. For the XSHOOTER observations, we find that a typical Lyman-$\alpha$ emission line with an intrinsic velocity width of about $200$\,km\,s$^{-1}$ would be robustly picked up in our spectroscopy if it represented a S/N of 3.5-4$\sigma$ (using a 5\AA\,$\times$0\farcs8 aperture for the Near-IR channel and a 4\AA\,$\times$0\farcs8 aperture for the optical channel).  In the case of our GNIRS observations, whose data reduction is subject to more systematics than our XSHOOTER observations, we impose a stricter 5$\sigma$ detection threshold for a line with intrinsic velocity of $200$\,km\,s$^{-1}$.  Some of the GNIRS systematics arise from the low-refractive index layers of the anti-reflection coatings on the GNIRS lenses which contain radioactive thorium, causing the array to be peppered with spikes during long exposures. Additionally, the
GNAAC controller of GNIRS superimposes systematics (vertical striping and horizontal banding).

\begin{figure}
  \resizebox{0.50\textwidth}{!}{\includegraphics{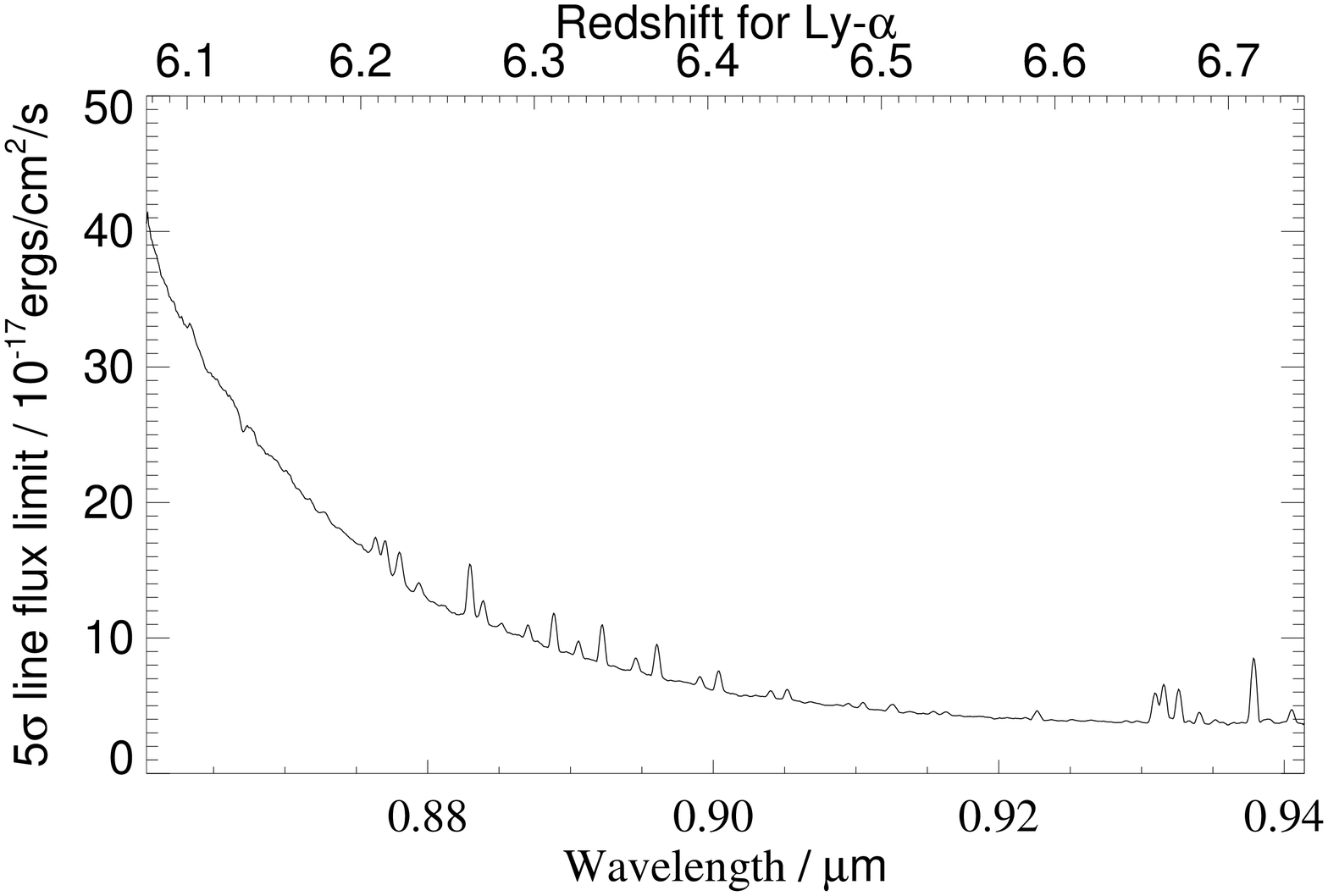}} \\
   \resizebox{0.50\textwidth}{!}{\includegraphics{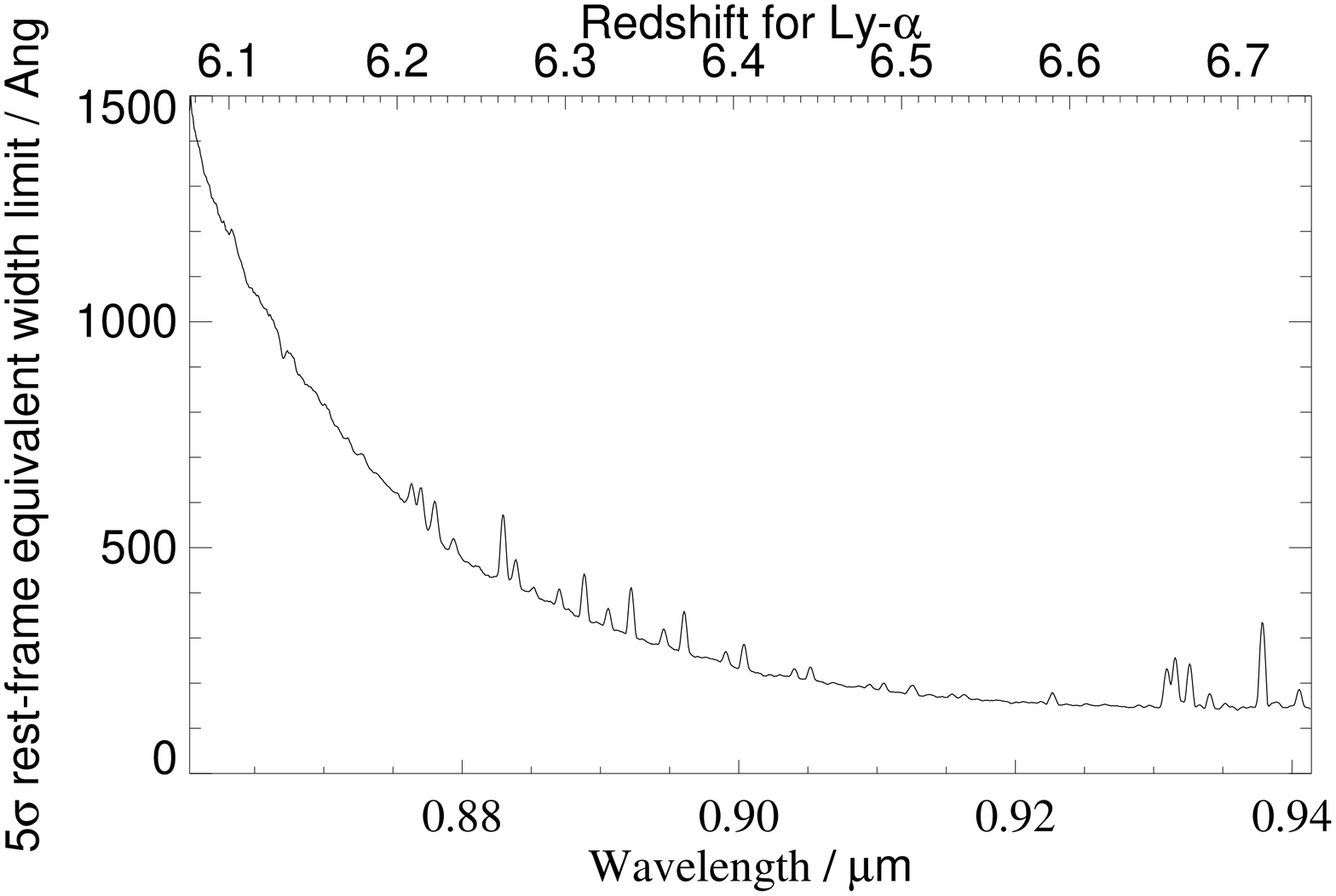}} \\
 \caption{Line flux and Equivalent Width limits for zD1 from GNIRS observations.  The upper panel shows the 5$\sigma$ line flux limit probed by Order 5 of our spectroscopy using a grating angle of 21 and the XD\_G0507 filter.  This filter does not have good transmission in the 0.8-1.2$\micron$ range, which is why the sensitivity is not very good.  The lower panel shows the 5$\sigma$ Equivalent Width limit for the redshift range probed by the same order of the same spectroscopic setting.  Y-mag: 26.71.  This plot (and all other EW upper-limit plots for GNIRS) assumes a spectrally unresolved source; a typical line with intrinsic velocity width of 200km$s^{-1}$ would have to be about 2 times brighter than an unresolved line to be robustly detected in our GNIRS spectroscopy.}
 \label{fig:EWlimitzD1}
\end{figure}

\begin{figure}
  \resizebox{0.50\textwidth}{!}{\includegraphics{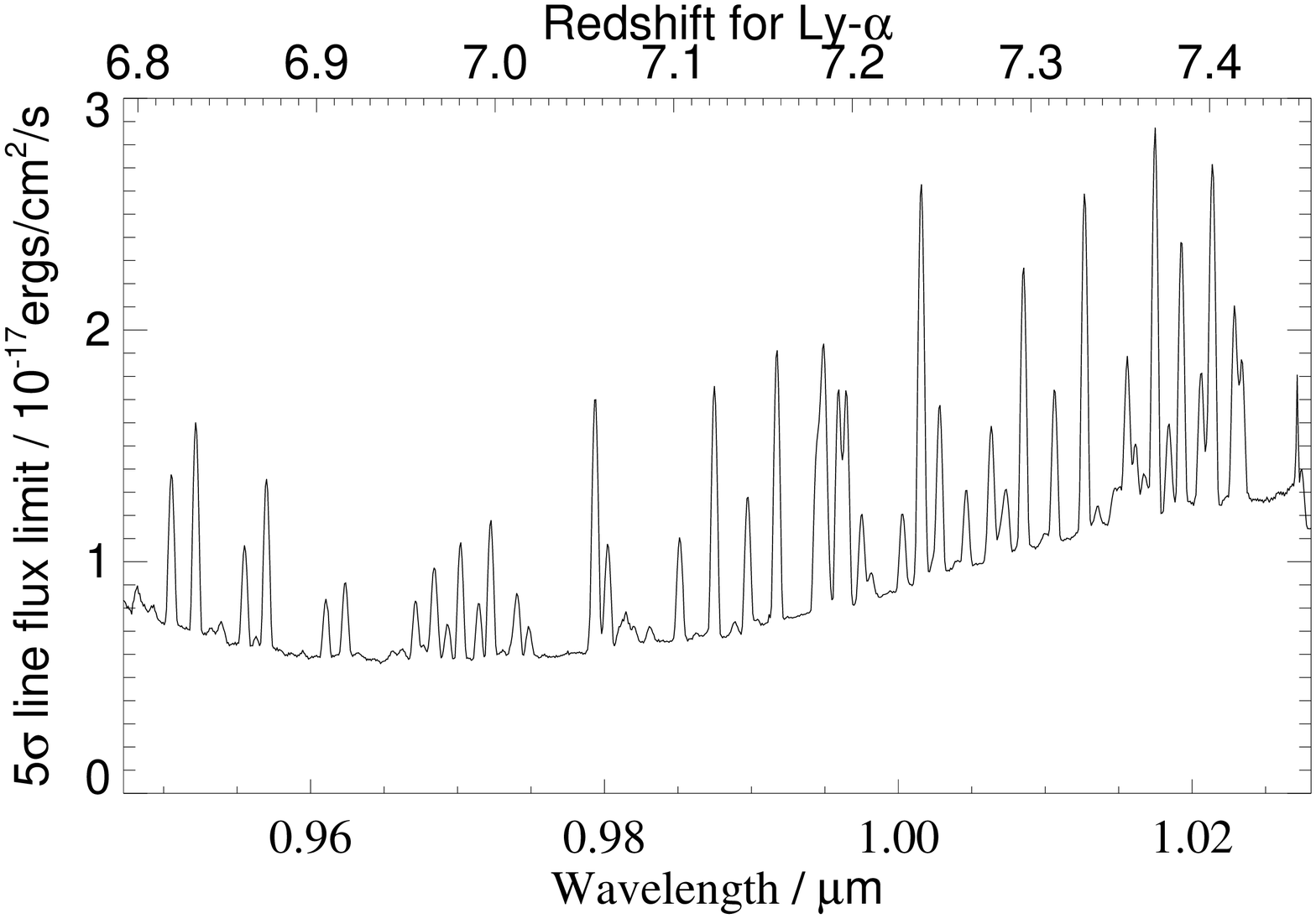}} \\
   \resizebox{0.50\textwidth}{!}{\includegraphics{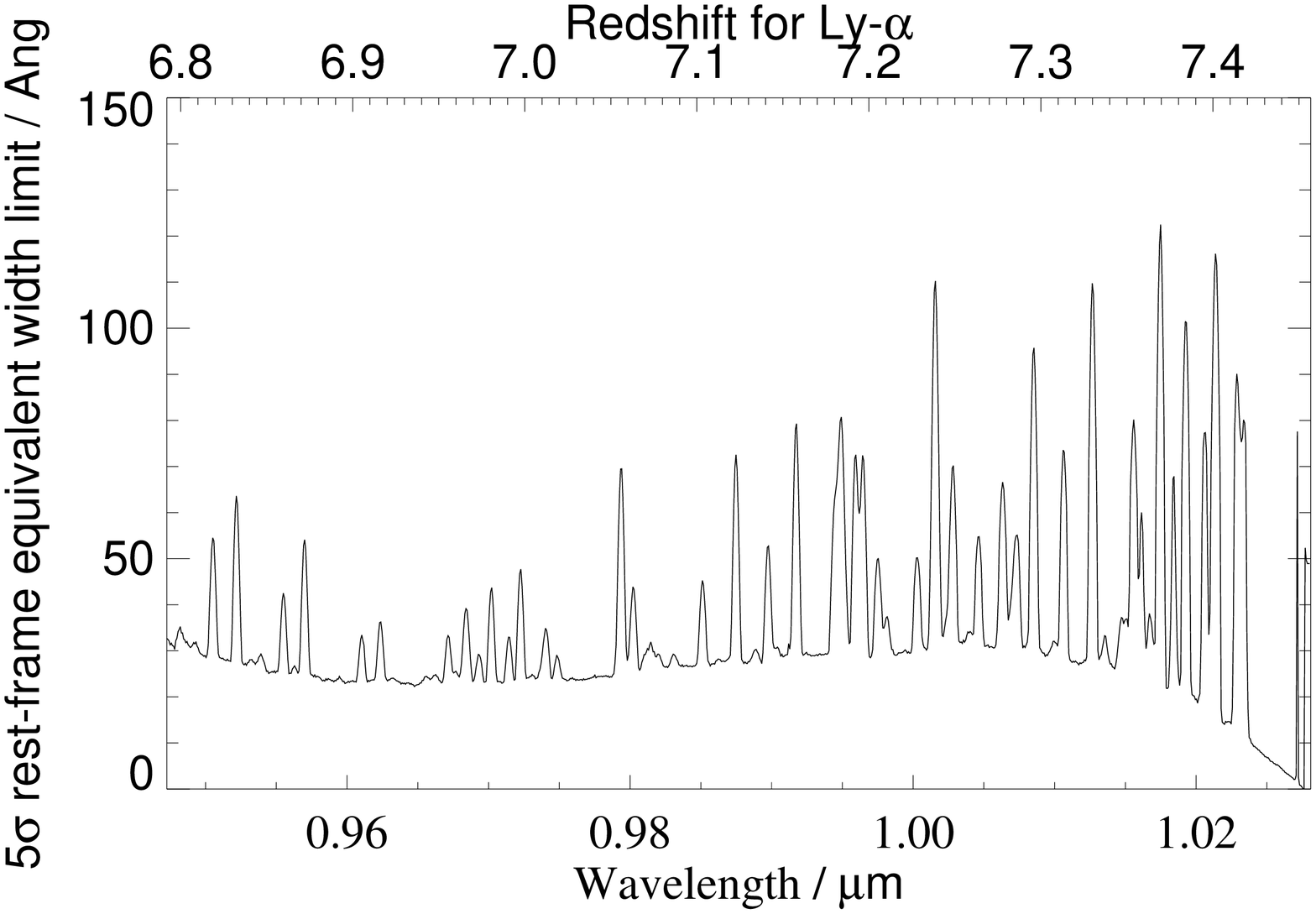}} \\
 \caption{Line flux and Equivalent Width limits for zD1 from GNIRS observations (continued).  The upper panel shows the 5$\sigma$ line flux limit probed by Order 5 of our spectroscopy using a grating angle of 23.2 and the the XD\_G0525 filter.  The lower panel shows the 5$\sigma$ Equivalent Width limit for the redshift range probed by the same order of the same spectroscopic setting.  Y-mag: 26.71.}
 \label{fig:EWlimitzD1b}
\end{figure}

\begin{figure}
  \resizebox{0.50\textwidth}{!}{\includegraphics{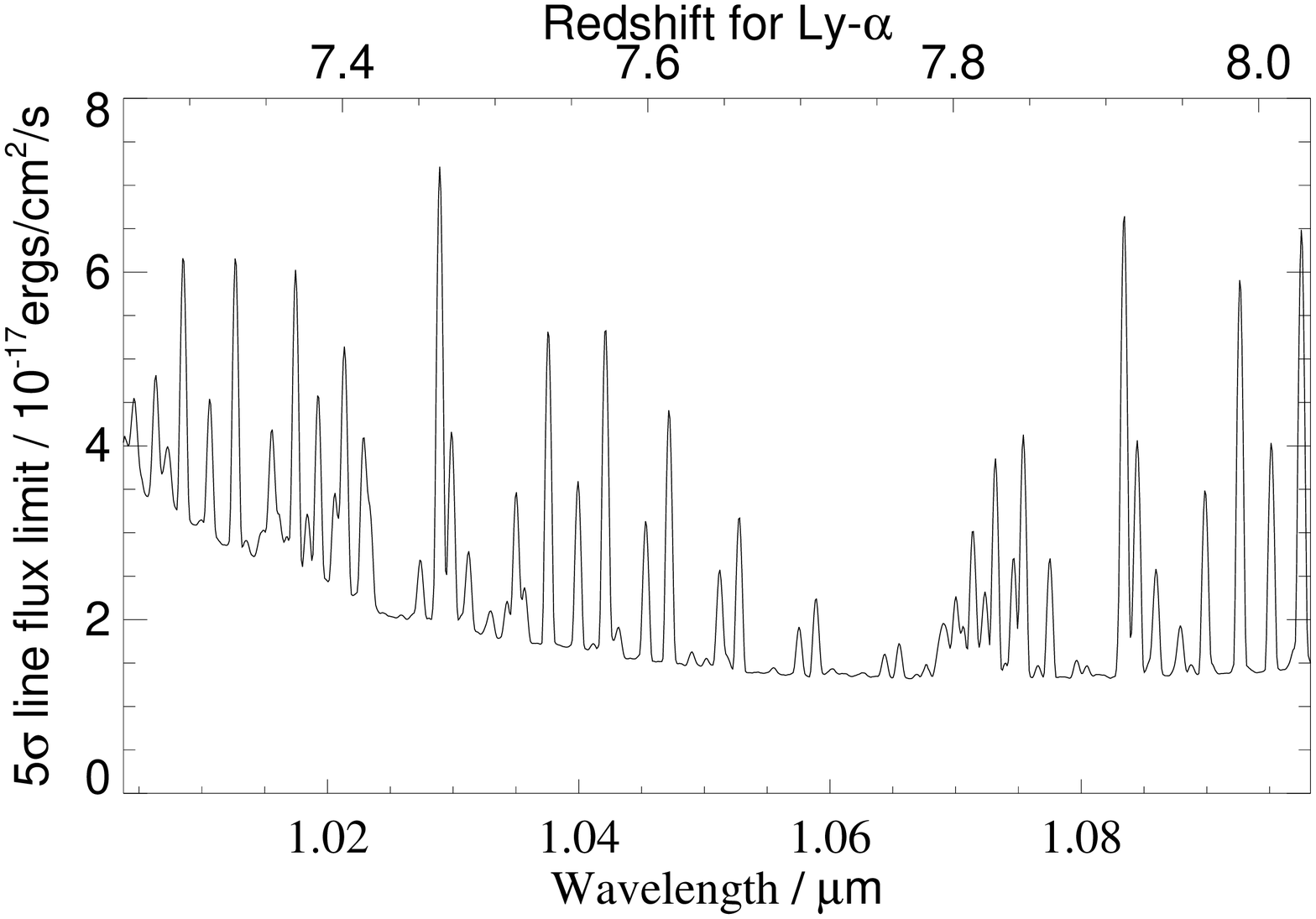}} \\
   \resizebox{0.50\textwidth}{!}{\includegraphics{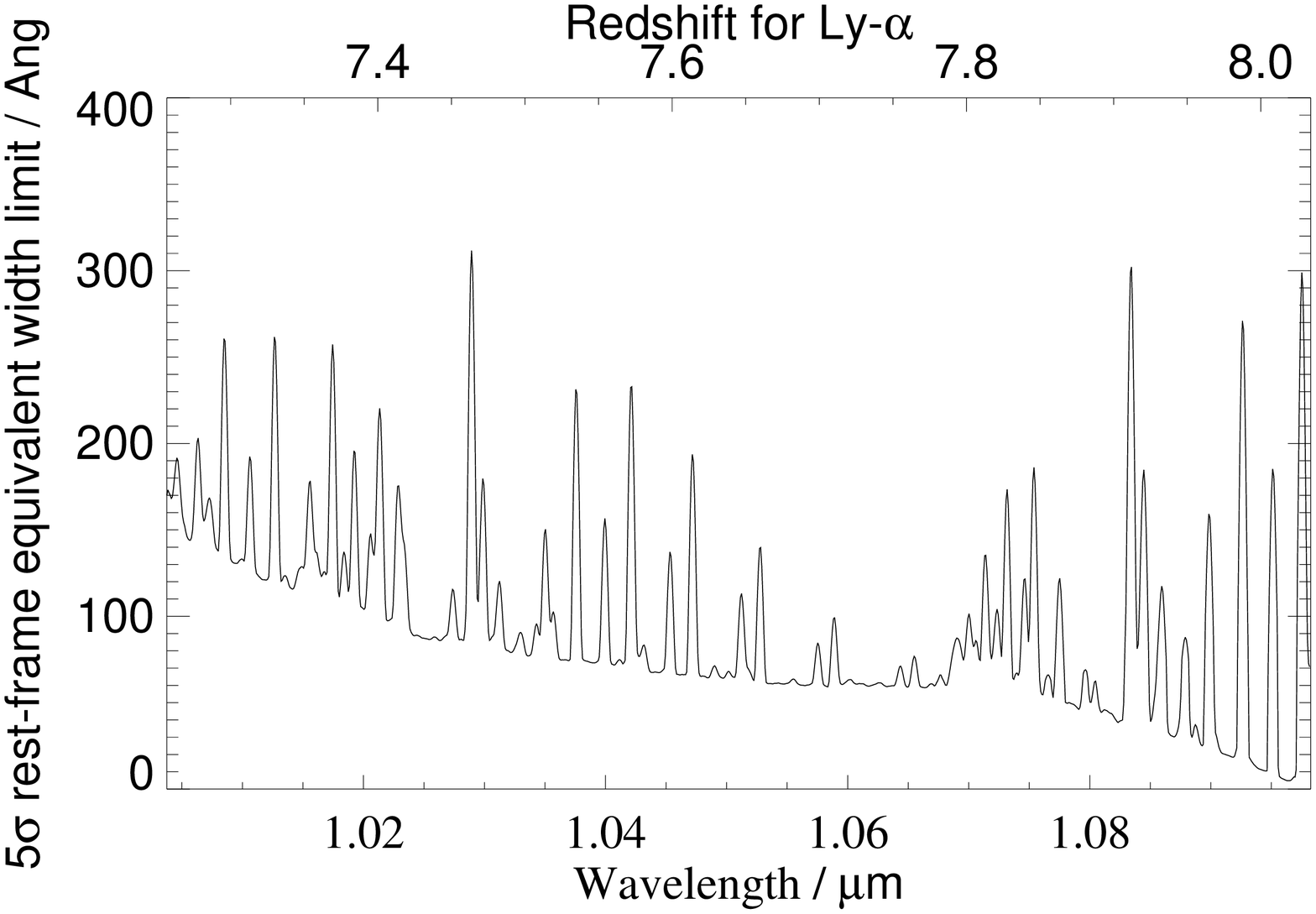}} \\
 \caption{Line flux and Equivalent Width limits for zD1 from GNIRS observations (continued).  The upper panel shows the 5$\sigma$ line flux limit probed by Order 4 of our spectroscopy using a grating angle of 21 and the XD\_G0507 filter.  The lower panel shows the 5$\sigma$ Equivalent Width limit for the redshift range probed by the same order of the same spectroscopic setting.  Y-mag: 26.71.}
 \label{fig:EWlimitzD1c}
\end{figure}

\begin{figure}
   \resizebox{0.50\textwidth}{!}{\includegraphics{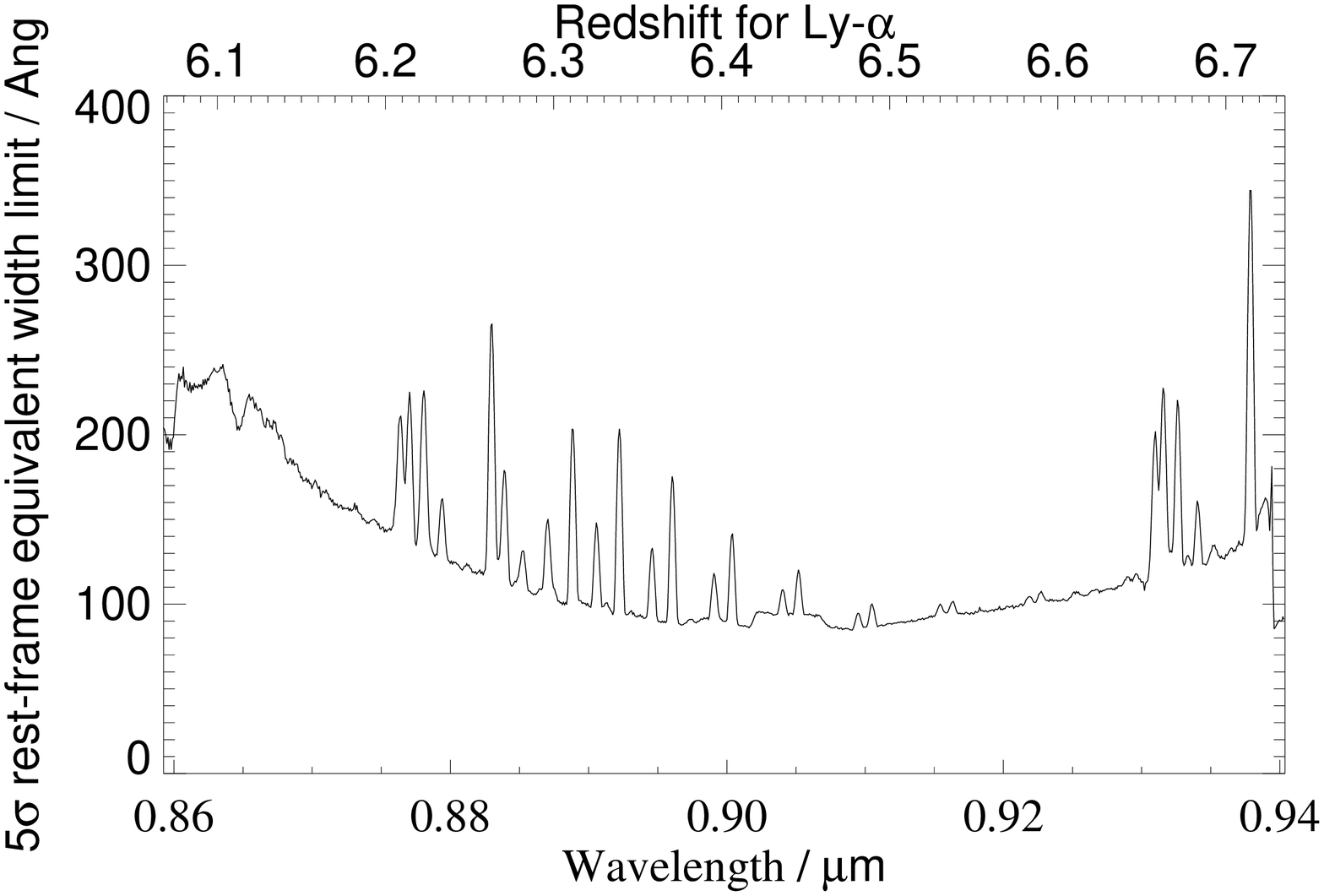}} \\
   \resizebox{0.50\textwidth}{!}{\includegraphics{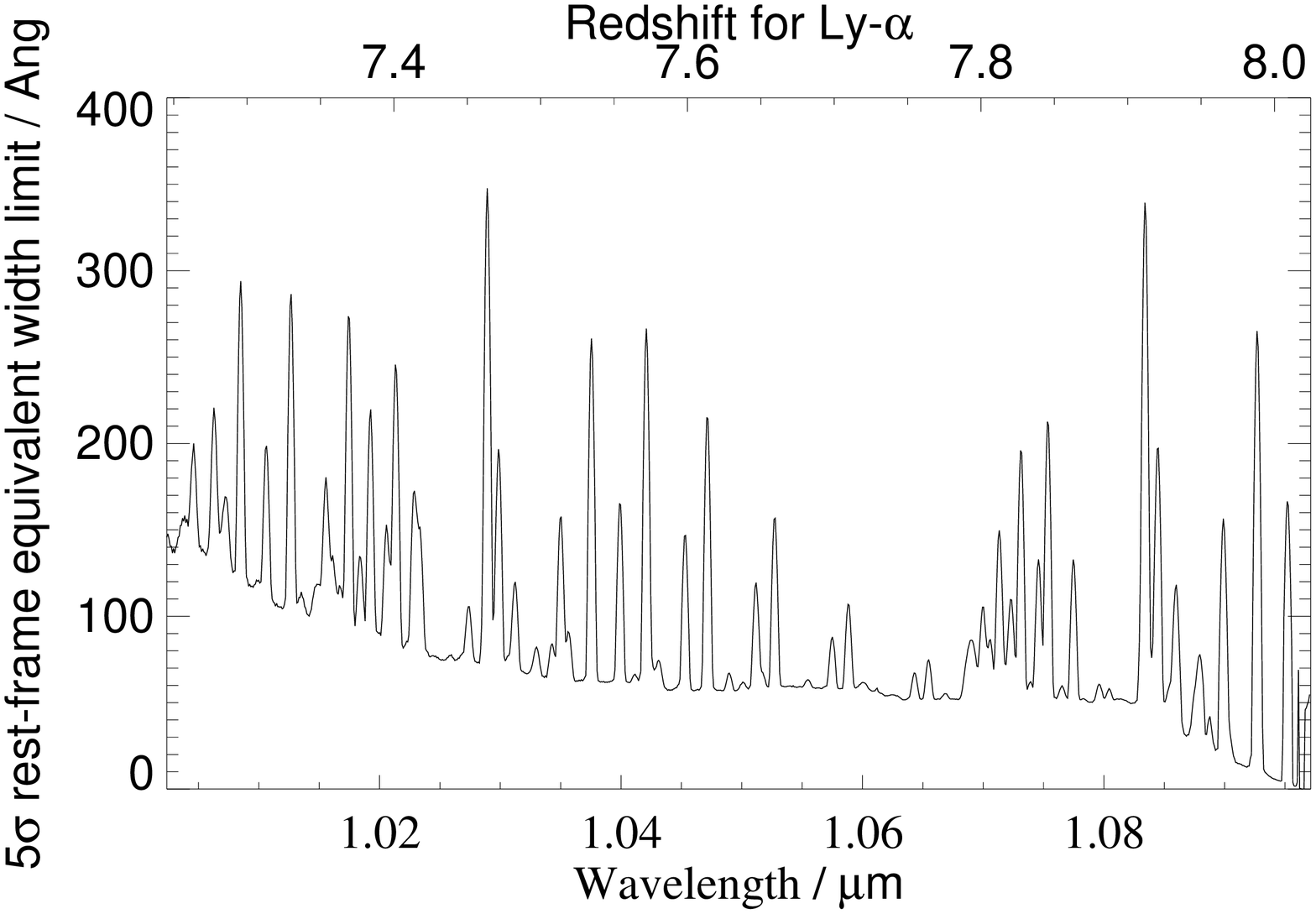}} \\
 \caption{5$\sigma$ Equivalent Width limit for zD2 from GNIRS observations.  The upper panel shows the redshift range probed by Order 5 of our spectroscopy using a grating angle of 21 and the XD\_G0525 filter.  The lower panel shows the redshift range probed by Order 4.  Y-mag: 27.48.} 
 \label{fig:EWlimitzD2}
\end{figure}

\begin{figure}
   \resizebox{0.50\textwidth}{!}{\includegraphics{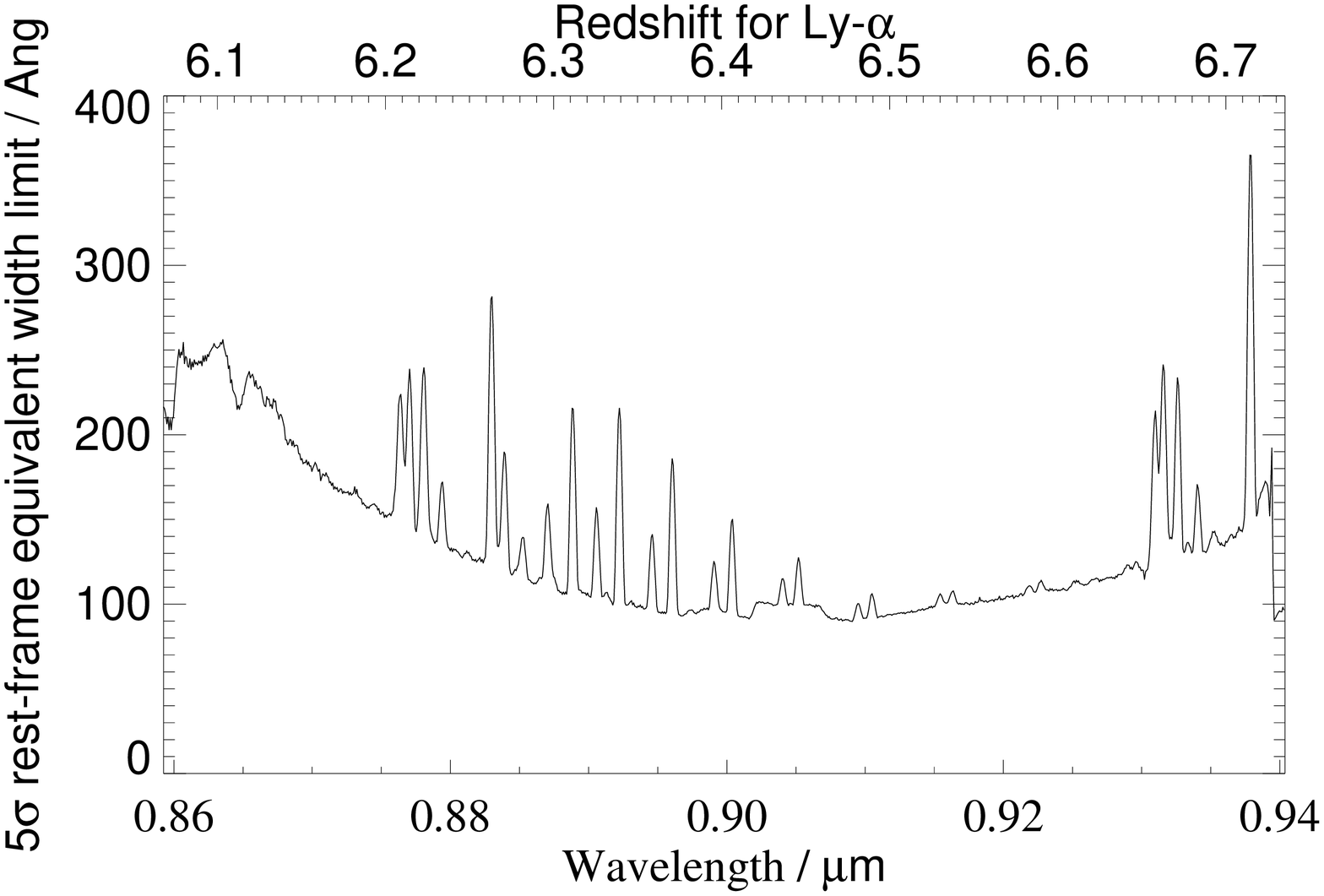}} \\
   \resizebox{0.50\textwidth}{!}{\includegraphics{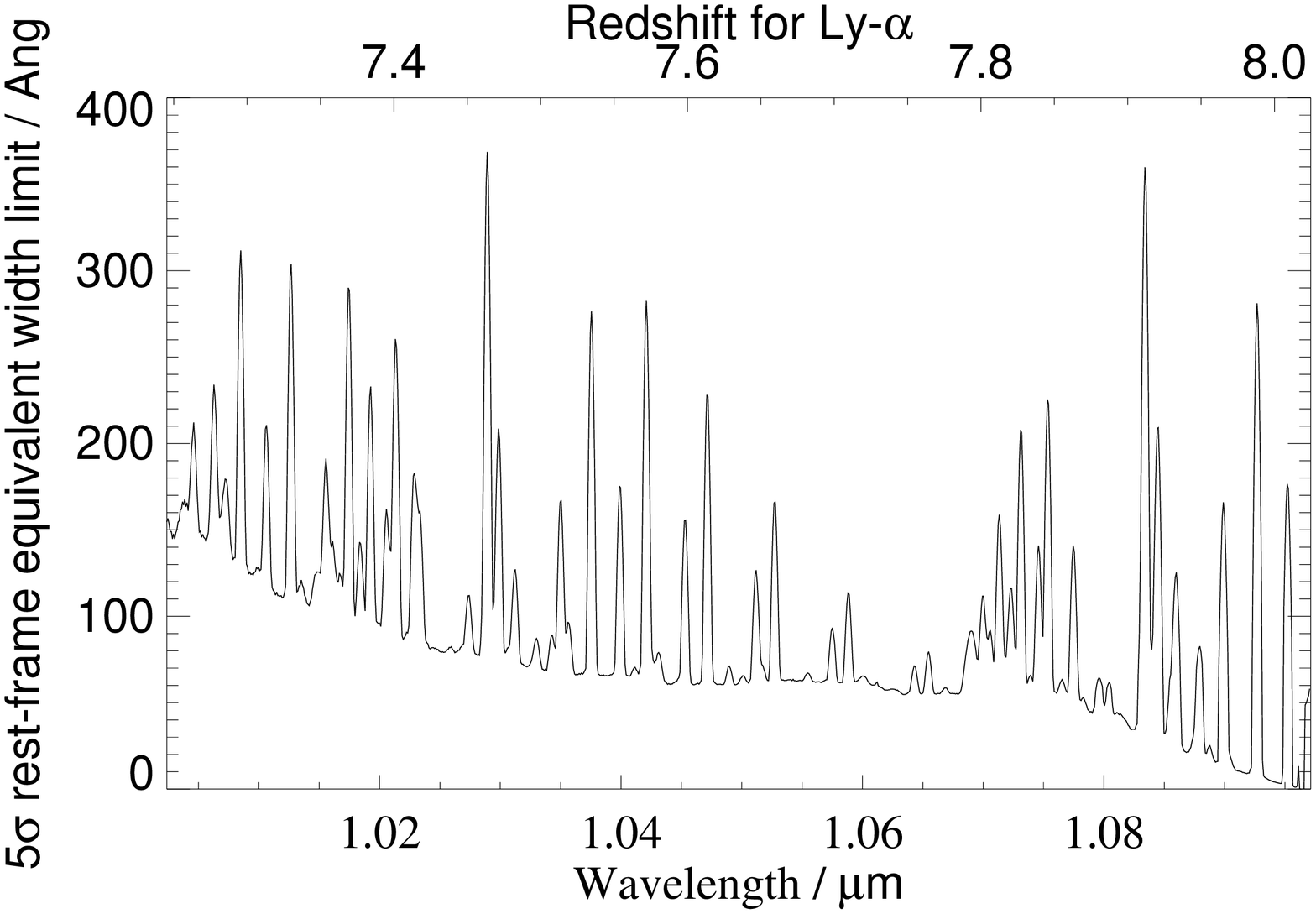}} \\
 \caption{5$\sigma$ Equivalent Width limit for zD3 from GNIRS observations.  The upper panel shows the redshift range probed by Order 5 of our spectroscopy using a grating angle of 21 and the XD\_G0525 filter.  The lower panel shows the redshift range probed by Order 4.  Y-mag: 27.5.} 
 \label{fig:EWlimitzD3}
\end{figure}

\begin{figure}
   \resizebox{0.50\textwidth}{!}{\includegraphics{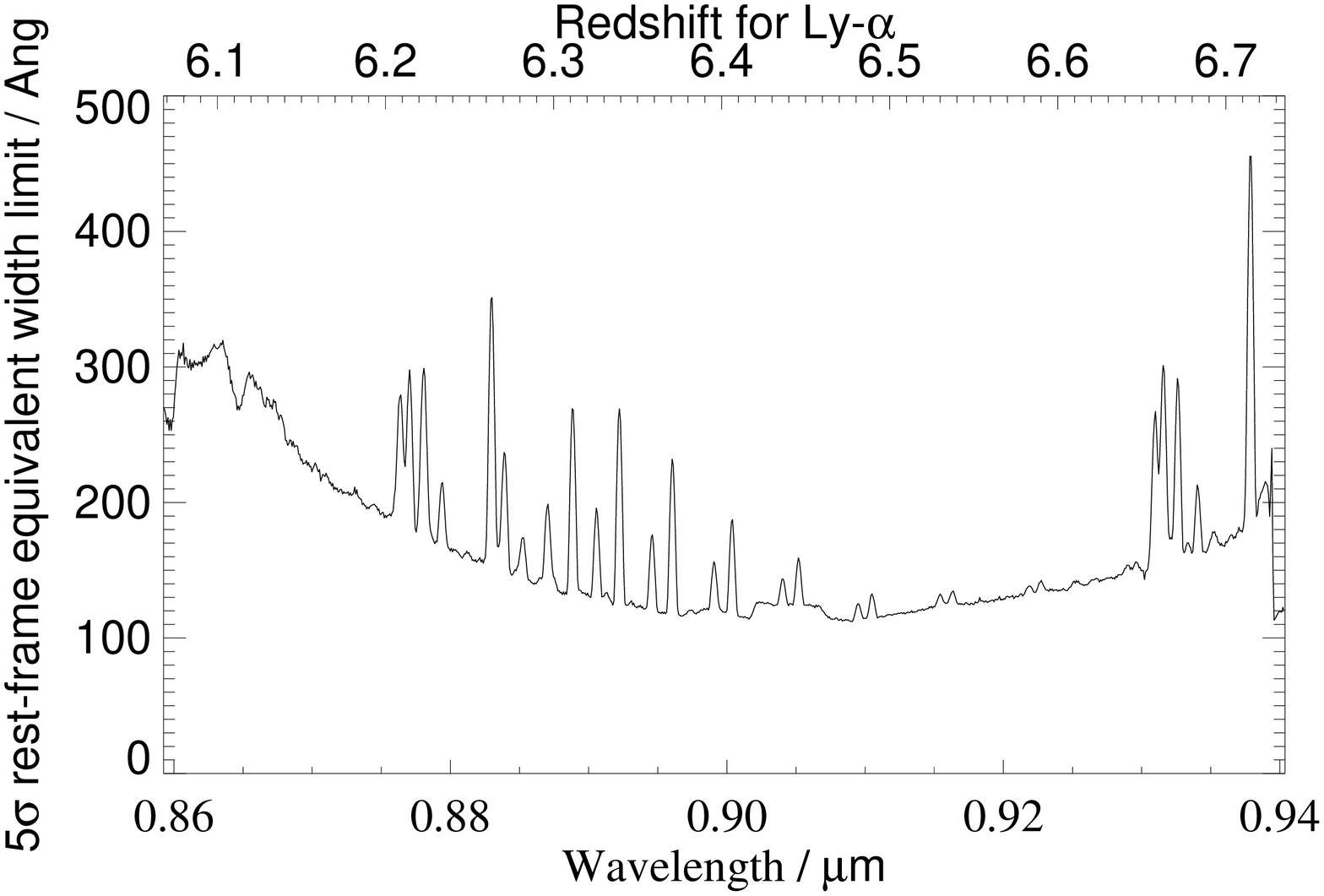}} \\
   \resizebox{0.50\textwidth}{!}{\includegraphics{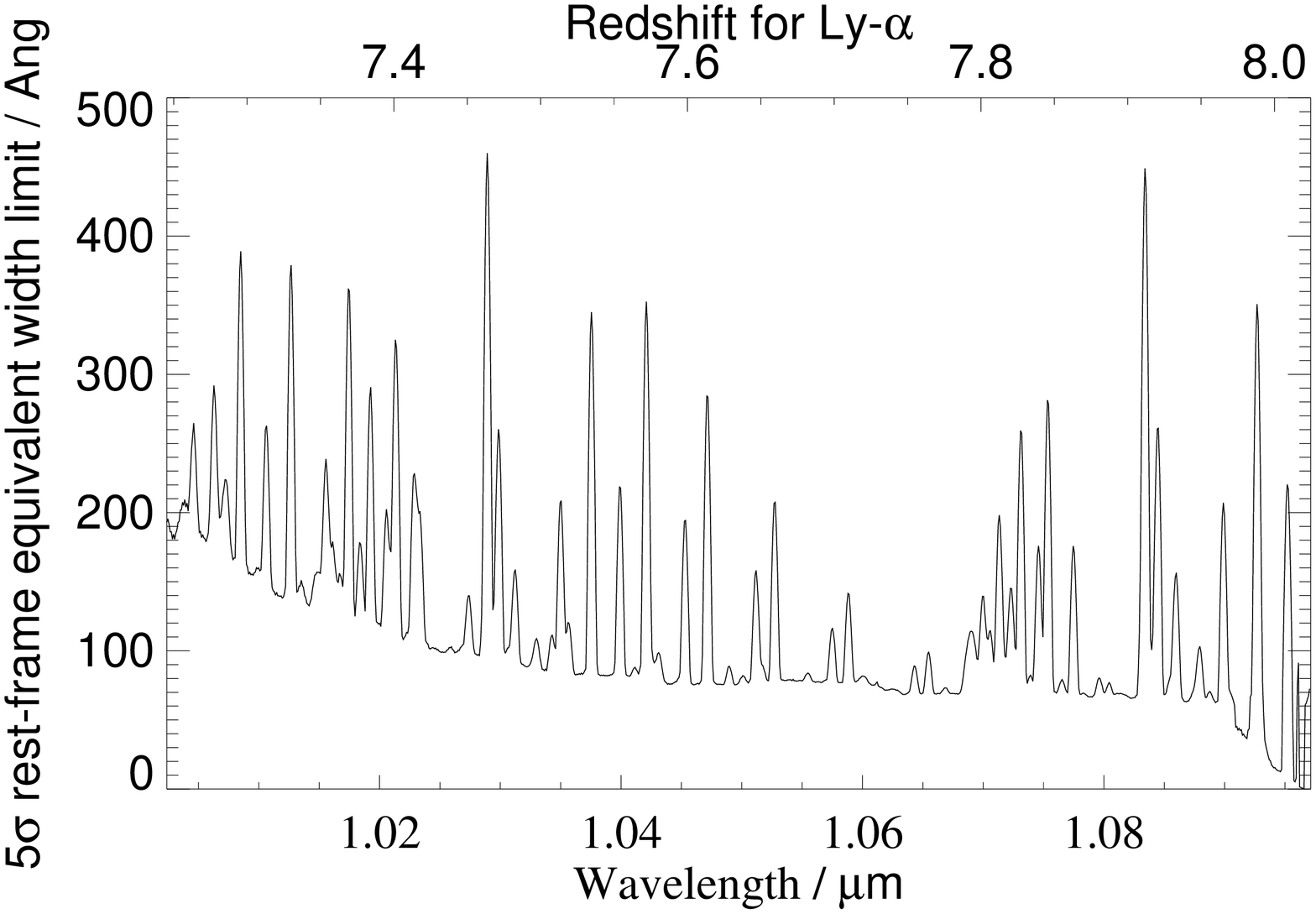}} \\
 \caption{5$\sigma$ Equivalent Width limit for zD4 from GNIRS observations.  The upper panel shows the redshift range probed by Order 5 of our spectroscopy using a grating angle of 21 and the XD\_G0525 filter.  The lower panel shows the redshift range probed by Order 4.  Y-mag: 27.84.} 
 \label{fig:EWlimitzD4}
\end{figure}

\begin{figure*}
   \resizebox{\textwidth}{!}{\includegraphics{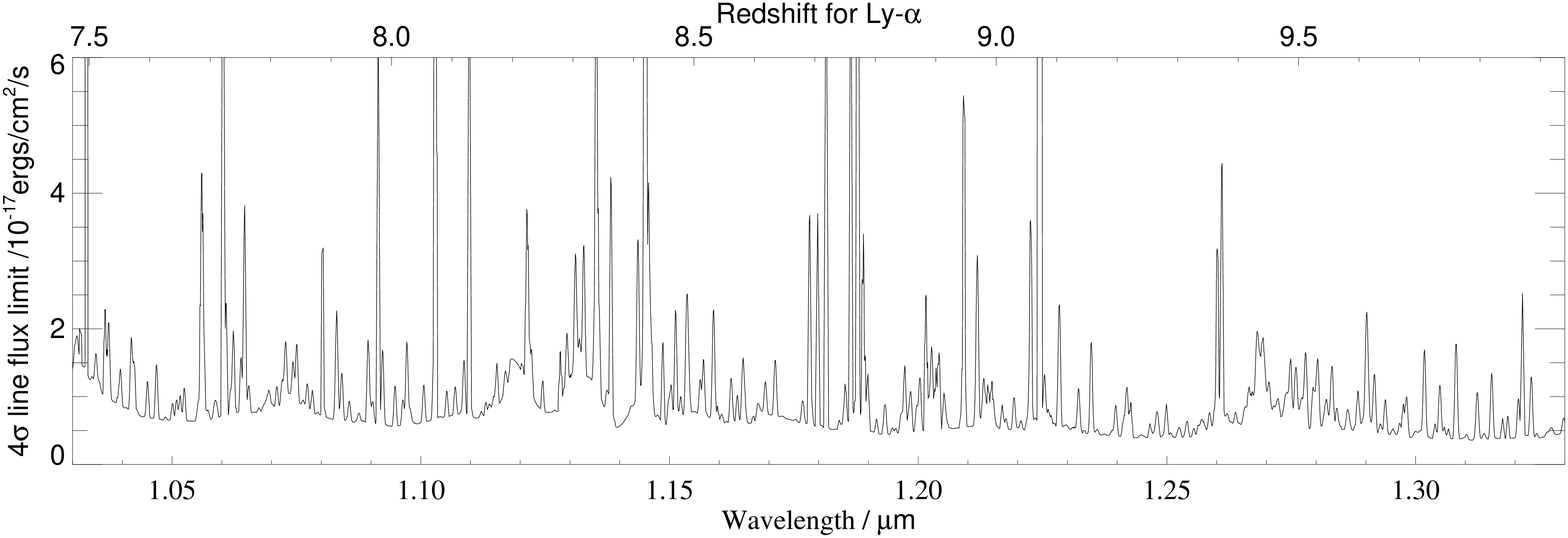}} \\
 \caption{4$\sigma$ line flux limit for our observations with the Near-IR channel of XSHOOTER.}
 \label{fig:IR_sensitivity}
\end{figure*}

\begin{figure*}
   \resizebox{\textwidth}{!}{\includegraphics{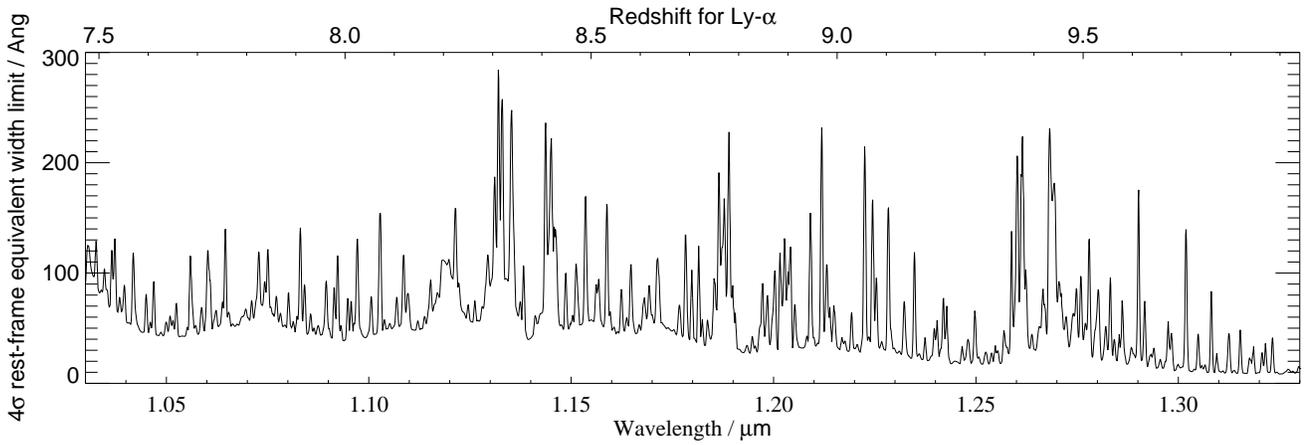}} \\
 \caption{4$\sigma$ Equivalent Width limit for ERS.YD2 from observations with XSHOOTER.  This plot and the next two EW upper-limit plots for XSHOOTER observations of HUDF.YD3 and P34.z.4809 assume a spectrally unresolved source; a typical line with intrinsic velocity width of 200km$s^{-1}$ would have to be about 1.5 times brighter than an unresolved line to be robustly detected in our XSHOOTER spectroscopy (i.e. the limits plotted here correspond to $\approx 3\sigma$ for such a spectrally resolved line).}
 \label{fig:ERSYD2}
\end{figure*}

\begin{figure*}
   \resizebox{\textwidth}{!}{\includegraphics{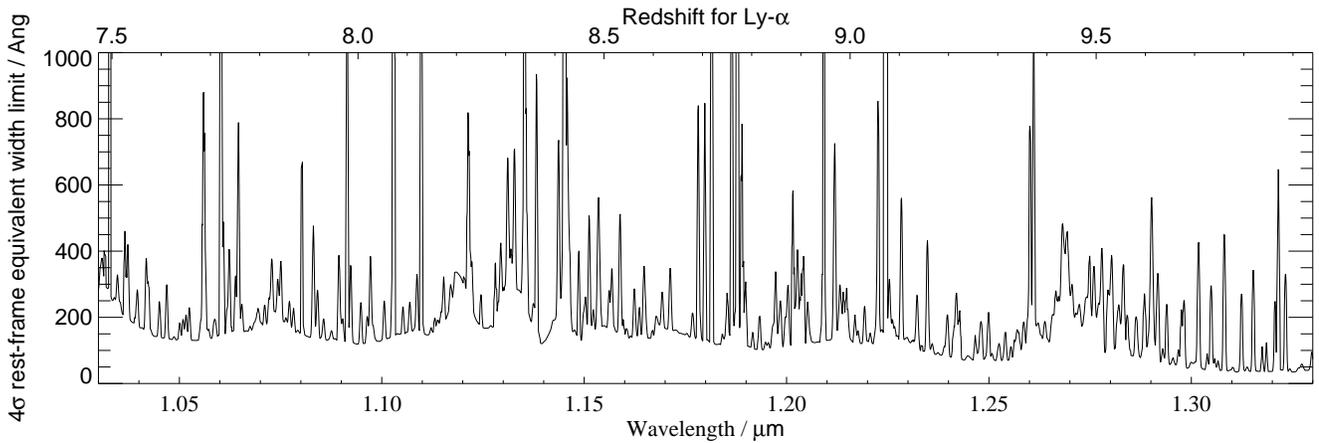}}
 \caption{4$\sigma$ Equivalent Width limit for HUDF.YD3 from observations with XSHOOTER.} 
 \label{fig:HUDFYD3J}
\end{figure*}

In conjunction with the continuum flux inferred from the {\em HST} imaging, our spectroscopy is deep enough to allow us to place interesting rest-frame equivalent-width (EW) limits on Lyman-$\alpha$ emission from our objects as shown in Figures \ref{fig:EWlimitzD1} -- \ref{fig:HUDFYD3J}.  The continuum is inferred from the broad-band photometry from HST, from the filter above the Lyman-$\alpha$ break, i.e.\ the F105W $Y$-band for the $z'$-drops at $z\approx 7$, and the F125W $J$-band for the $Y$-drops at $z\approx 8-9$. We assume here a rest-frame spectral slope above the Lyman-$\alpha$ break of the form $f_{\lambda}\propto \lambda^{beta}$, where we adopt $\beta=-2$ (equivalent to a spectrum flat in $f_{\nu}$), as found to be typical of $z>6$ galaxies (Stanway et al.\ 2005; Wilkins et al.\ 2011), including the galaxies surveyed here (Bunker et al.\ 2010). We experimented with a range of spectral slopes between a redder slope of $\beta =-1.5$ and a bluer slope of $\beta=-2.5$, and found that this only changed the inferred equivalent width limits by about 5 per cent (because the filter used to infer the continuum lies at wavelengths just beyond Lyman-$\alpha$). At the high redshift extreme of the expected redshift distribution (Figure~\ref{fig:sims}), the Lyman-$\alpha$ break can encroach on the shortest wavelengths of the {\em HST} bandpass used to infer the continuum, and we correct for this assuming total absorption below the Lyman-$\alpha$ wavelength at these redshifts (the Gunn-Peterson effect). In these cases we also correct for possible Lyman-$\alpha$ line emission contamination of the broad-band magnitude, given our upper limits on the line flux. The significance of the upper limits on the flux of an emission line, given our non-detections, depends on the wavelength (because the sky spectrum, atmospheric transmission and detector sensitivity vary with wavelength), and also on the spatial and spectral extent of any line emission.

The Lyman break galaxies at $z>6$ are typically very compact in {\em HST} images (e.g., Oesch et al.\ 2010b finds $r\approx 0.7$\,kpc) and are unresolved in ground-based seeing. Hence we adopt a spatial extraction aperture about 1.5 times the seeing disk in order to maximize the signal-to-noise ratio. The spectral extent (i.e.\ velocity width) of the Lyman-$\alpha$ emission is less certain. We consider two scenarios: one where the line emission is unresolved at the spectral resolution of GNIRS or XSHOOTER (i.e.\ $\Delta v_{\rm FWHM}<100$\,km\,s$^{-1}$), and the other where the intrinsic line width is around $200$\,km\,s$^{-1}$, similar to that seen in some Lyman break galaxies at $z\approx 6$ which have Lyman-$\alpha$ emission (e.g.\ Bunker et al.\ 2003, Stanway et al.\ 2004b).  

For XSHOOTER, our spectral resolution is $\approx 2.5$\,\AA\ FWHM in the near-infrared (the optical arm of XSHOOTER has higher resolution of 1.4\,\AA ), and a line with intrinsic velocity width of $\Delta v_{\rm FWHM}=200$\,km\,s$^{-1}$ would result in observed line widths of $6.5-8$\,\AA\ FWHM, after convolution with the spectral resolution of the instruments. Hence we adopt a spectral extraction width of 4\,\AA\, (10 pixels) in the optical and 5\,\AA\, (5 pixels) for the infrared channels of XSHOOTER, intermediate in size between the wavelength spread of the emission lines in our two scenarios.  Spatially, we adopt a size of $0\farcs8$ for our aperture (which is 5pix in the optical channel and 4pix in the infrared channel).  In the case of an unresolved line, we capture $95.4\%$ of the flux with our aperture in our optical data and $87\%$ of the flux in our Near-Infrared data.  For a $200$\,km\,s$^{-1}$ line we capture $48.5\%$ of the flux with our aperture in our optical data and $52.6\%$ of the flux in our Near-Infrared data.  We apply the above aperture corrections in computing the flux limits.

In the case of the GNIRS observations, the spectral resolution is $\approx 2.56$\,\AA\ FWHM, and a line with an intrinsic velocity width of $\Delta v_{\rm FWHM}=200$\,km\,s$^{-1}$ would correspond to an observed line width of $\approx7$\,\AA\ FWHM after convolution with the spectral resolution of the instrument.  We adopt a spectral extent of $3.96$\,\AA\ for our aperture (corresponding to 5 pixels in our pipeline output) and a spatial extraction width of $0\farcs75$ (5pixels).  Our square aperture ($5\times5$-pixels) would capture $88\%$ of the flux from a spectrally unresolved object, whereas for a $200$kms$^{-1}$ it would capture 45\% of the flux.

\begin{figure*}
   \resizebox{\textwidth}{!}{\includegraphics{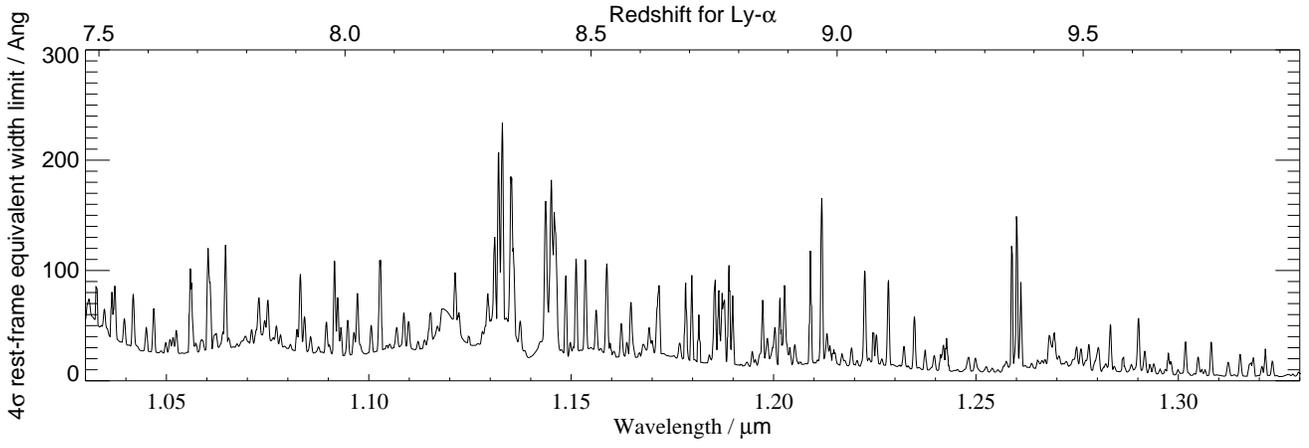}}
 \caption{4$\sigma$ Equivalent Width limit for UDF092y-03751196d from observations with XSHOOTER.} 
 \label{fig:UDF092y-03751196d}
\end{figure*}

\begin{figure*}
   \resizebox{\textwidth}{!}{\includegraphics{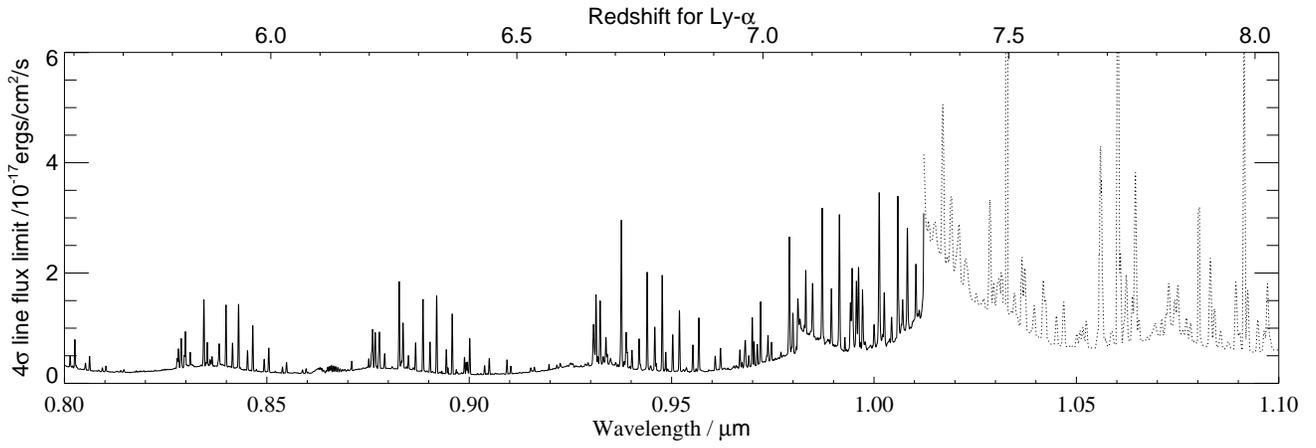}} \\
 \caption{4$\sigma$ line flux limit for our observations of P34.z.4809 with the Optical and Near-IR channels of XSHOOTER.  The region of the figure shown with a dotted line represents the limit obtained form data acquired with the NIR channel.}
 \label{fig:sensitivity_nir_opt}
\end{figure*}

\begin{figure*}
   \resizebox{\textwidth}{!}{\includegraphics{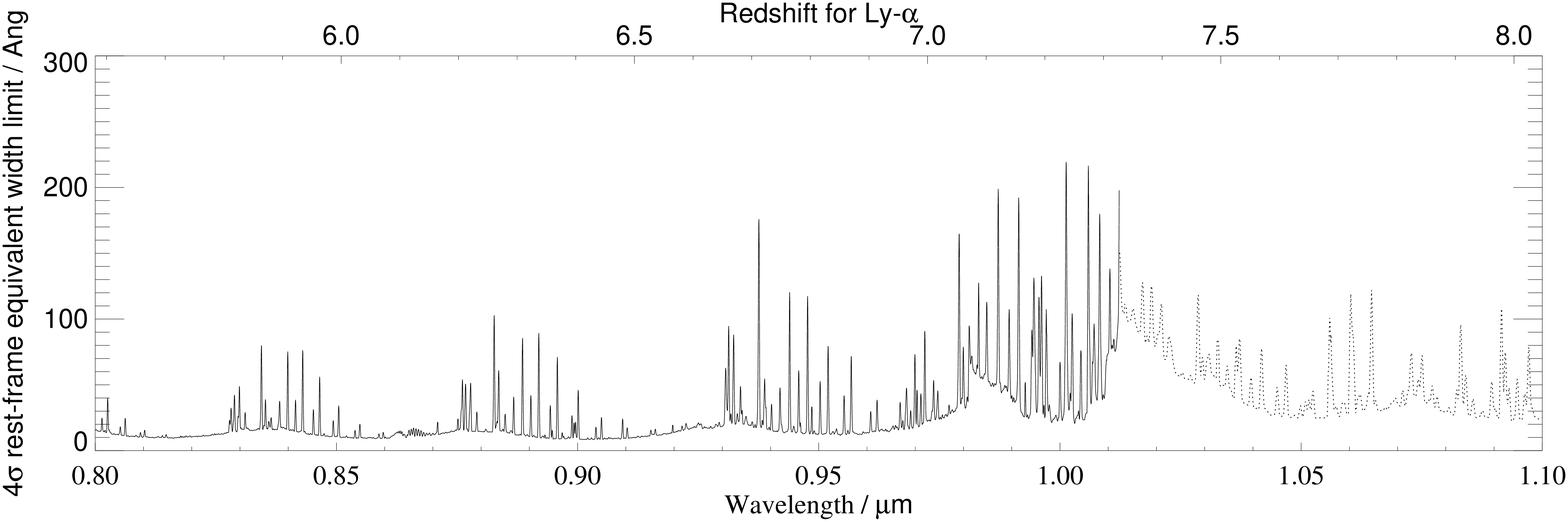}}
 \caption{4$\sigma$ Equivalent Width limit for P34.z.4809 from observations with XSHOOTER.  The region of the figure shown with a dotted line represents the limit obtained from data acquired with the NIR channel.} 
 \label{fig:P34Z4809}
\end{figure*}

\begin{figure}
   \resizebox{0.50\textwidth}{!}{\includegraphics{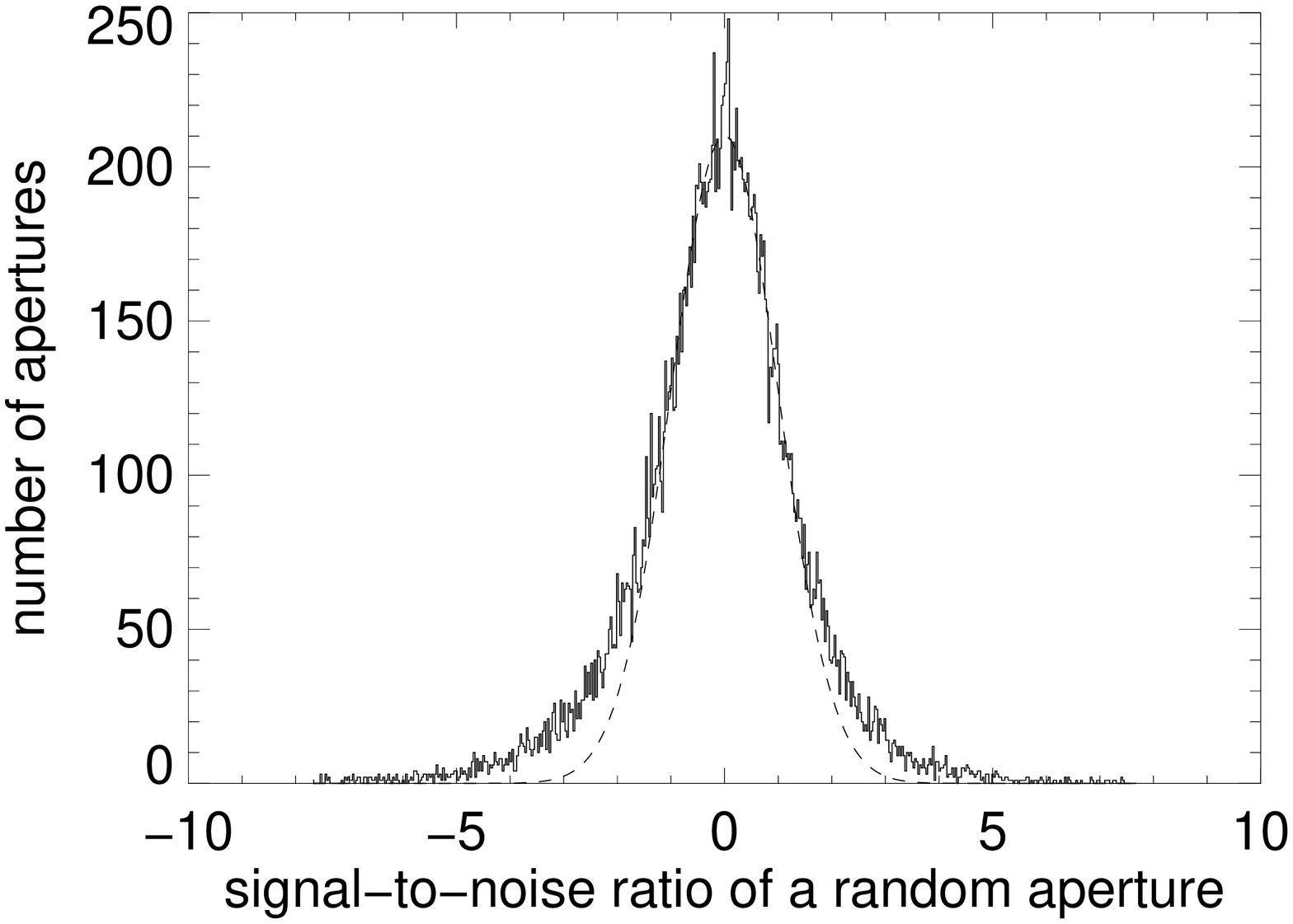}} 
 \caption{This plot for one of our XSHOOTER Near-InfraRed spectra shows the histogram of measured counts in independent $5\times 4$ pixel apertures.  The core distribution is well-fit by a Gaussian with the expected noise properties. The non-Gaussian extremes of the distribution are due to a small number of data reduction artefacts.} 
 \label{fig:sigma}
\end{figure}

We checked the validity of our 2D noise model by placing the chosen extraction apertures at random
on the 2D spectrum normalised by the noise model, and fitting a Gaussian to a histogram of
the measured fluxes within the apertures (Figure~\ref{fig:sigma} shows the results of this
for all spatially-independent apertures in one of our XSHOOTER near-infrared spectra).  The noise distribution was well-fit by a Gaussian with the expected noise properties. Some excess power in the wings (both positive and negative) was attributable to occasional isolated hot pixels or cosmic rays which had survived clipping in the data reduction, or sky line subtraction residuals. All $>4\,\sigma$ events (measured in the adopted apertures) were investigated and none were found to be consistent with emission lines.

\begin{figure*}
  \resizebox{0.85\textwidth}{!}{\includegraphics{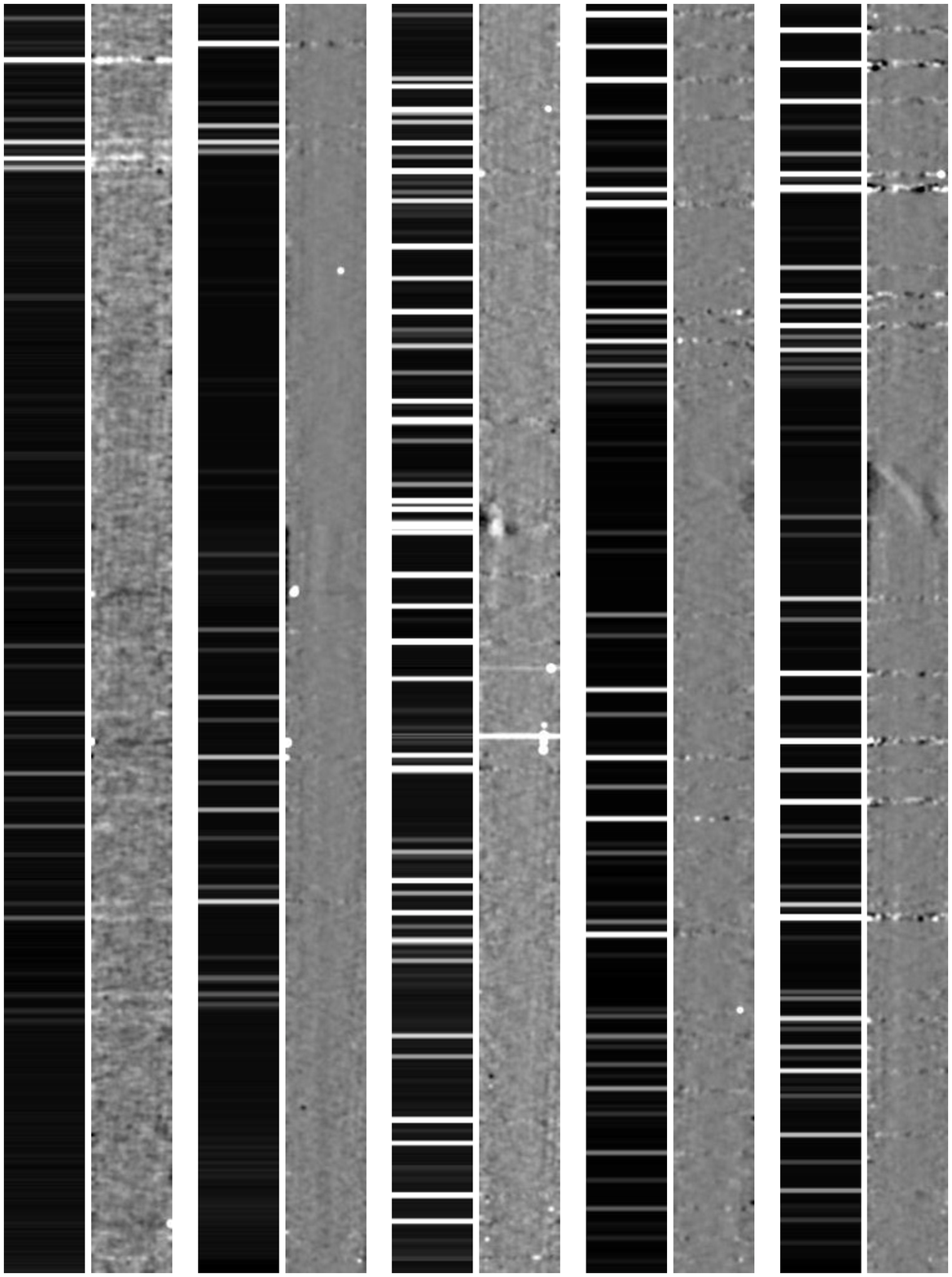}}
  \caption{This figure shows typical reduced 2D frames from all the different spectroscopic settings used for our GNIRS observations, together with the corresponding 2D frames showing sky emission lines.  From top to bottom, (a) the 0.84-0.94$\mu$m wavelength range covered with the XD\_G0507 filter and a grating angle of 21 (order 5), (b) the same wavelength region covered with the XD\_G0525 filter and a grating angle of 21 (order 5), (c) the 0.95-1.03$\mu$m wavelength range covered with the G\_0525 filter and a grating angle of 23.2 (order 5), (d) the 1.005-1.095$\mu$m wavelength range covered with the G\_0507 filter and a grating angle of 21 (order 4), and (e) the same wavelength region covered with the XD\_G0525 filter and a grating angle of 21 (order 4).  Wavelength increases from bottom to top.}
\label{fig:2dplots}
\end{figure*}

\begin{figure*}
	\resizebox{0.9\textwidth}{!}{\includegraphics{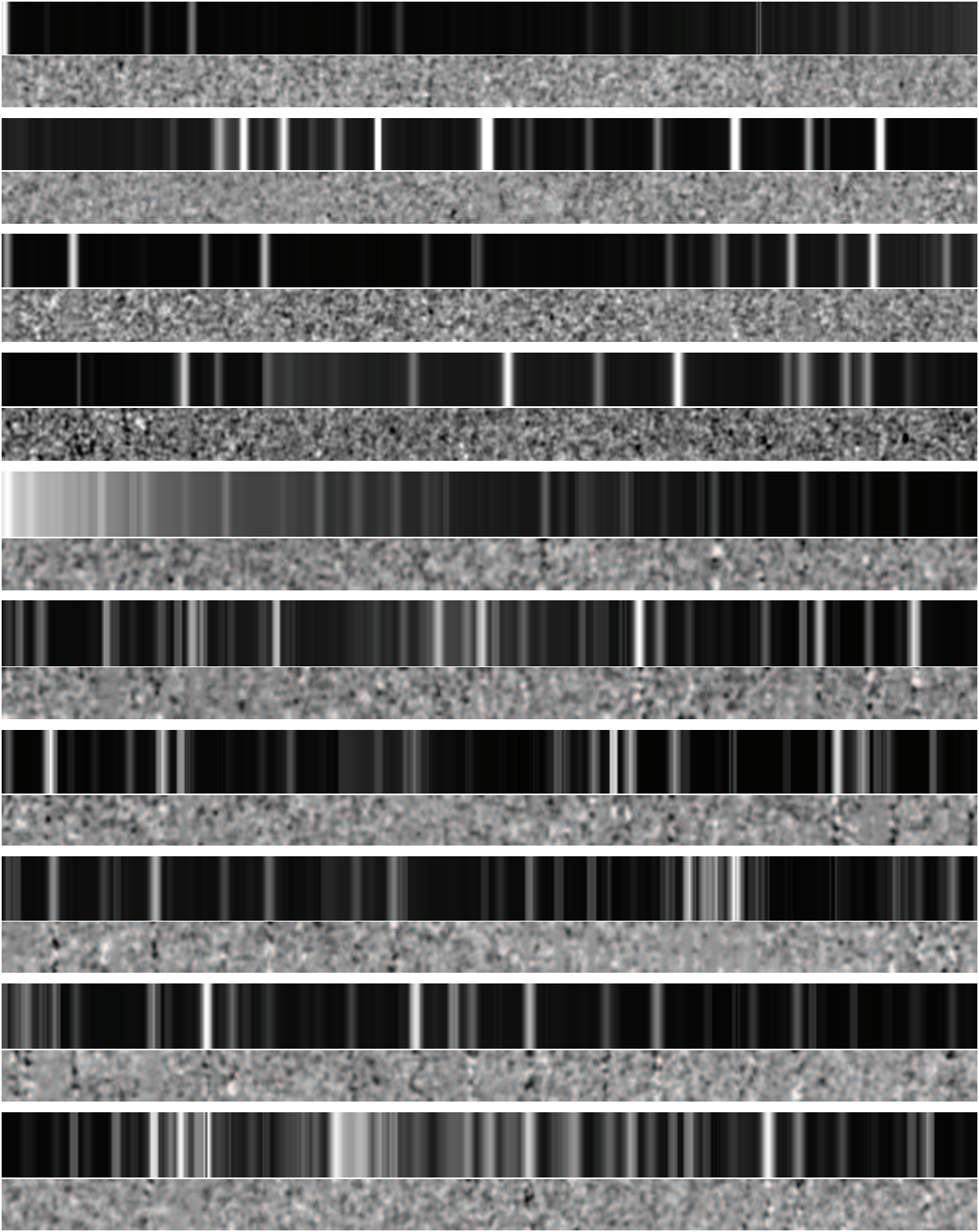}}
\caption{This figure shows typical reduced 2D frames for XSHOOTER, together with the corresponding 2D frames showing sky emission lines.  Displayed here is the wavelength range where Lyman-$\alpha$ is expected to lie for $z$-drops and $y$-drops.  From top to bottom, the first four panels show reduced data from the optical channel (here showing P34.z.4809), and together span the 0.9$\mu$m-1.0$\mu$m wavelength range, each panel spanning 0.025$\mu$m with wavelength increasing from left to right.  The following six panels show reduced data from the Near-IR channel (here showing UDF092y-03751196d) and together span the 1.0$\mu$m-1.3$\mu$m wavelength range, each panel spanning 0.05$\mu$m with wavelength increasing from left to right.}
\label{fig:2dplotsxshooter}
\end{figure*}

One of the objects targeted by our GNIRS spectroscopy, HUDF.zD1, has previousy been investigated by Fontana et al.\ (2010, their source G2\_1408) and they observe a tentative Lyman-$\alpha$ emission line at 9691.5\AA\ with $f_{Ly\alpha }=3.4\times 10^{-18}\,{\rm erg\,cm^{-2}\,s^{-1}}$.  We looked closely at this wavelength region in our GNIRS spectrum of this object.
At this wavelength, our measured 1\,$\sigma$ noise using a 3.96\AA$\times0\farcs75$ aperture (5$\times$5 pixels) is $1.01\times10^{-18}\,{\rm erg\,cm^{-2}\,s^{-1}}$. For a spectrally-unresolved emission line with the same flux as the Fontana et al.\ object, we would expect a $3\,\sigma$ detection in our GNIRS data. However, Fontana et al.\ report a marginally spectrally-resolved line profile of 10\,\AA\ FWHM (with an instrumental width of 7\,\AA\ FWHM for FORS\,2), corresponding to an intrinsic velocity spread of 200\,km\,s$^{-1}$ FWHM, after deconvolution with their instrumental resolution. At our GNIRS resolution we would expect such a line to have 7\,\AA\ FWHM. For this velocity profile and line flux, we would only expect a $1.7\,\sigma$ signal within our $5\times 5$\,pixel aperture. Hence our non-detection does not rule out the Fontana et al.\ line detection, particularly if it has significant velocity extent.  
We note that although HUDF.zD1 is well resolved in the WFC\,3 imaging and indeed comprises two  distinct components separated by 2\,kpc ($0\farcs37$) and with half-light radii of $0.5-0.8$\,kpc (Oesch et al.\ 2010b), in ground-based seeing this galaxy pair is unresolved. We have included the effects of the finite source size (before seeing) in our calculation of the slit losses.
We are currently analysing recently obtained VLT/FORS2 data, which will shed further light on this interesting target.

\begin{figure}
  \resizebox{0.47\textwidth}{!}{\includegraphics{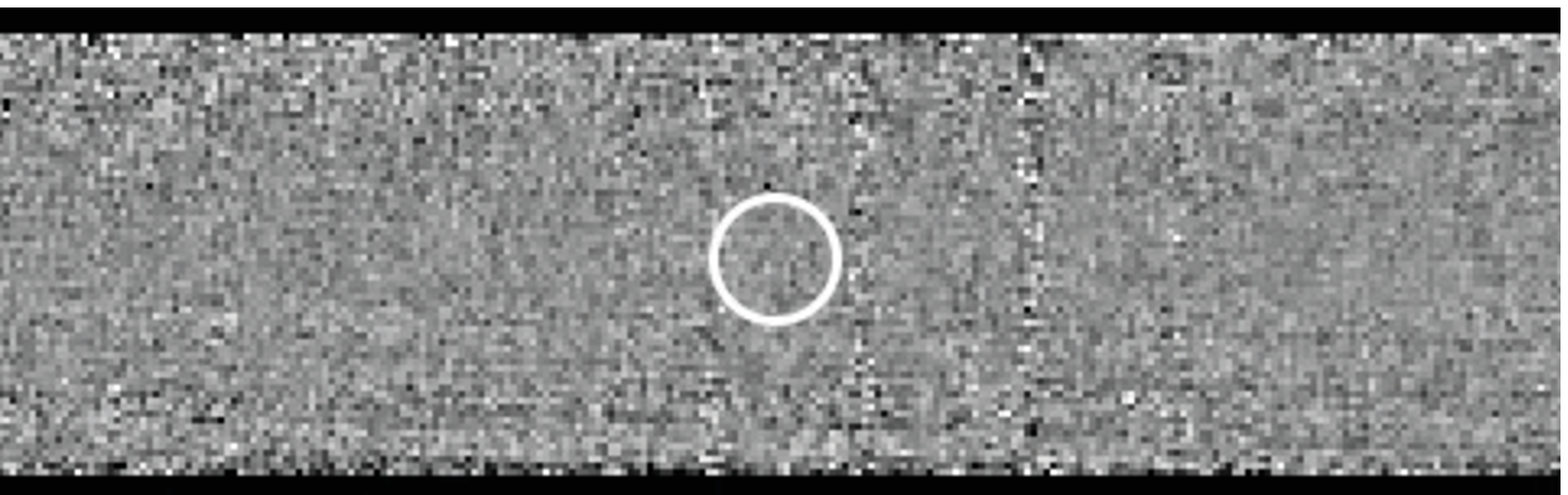}} \\
  \resizebox{0.47\textwidth}{!}{\includegraphics{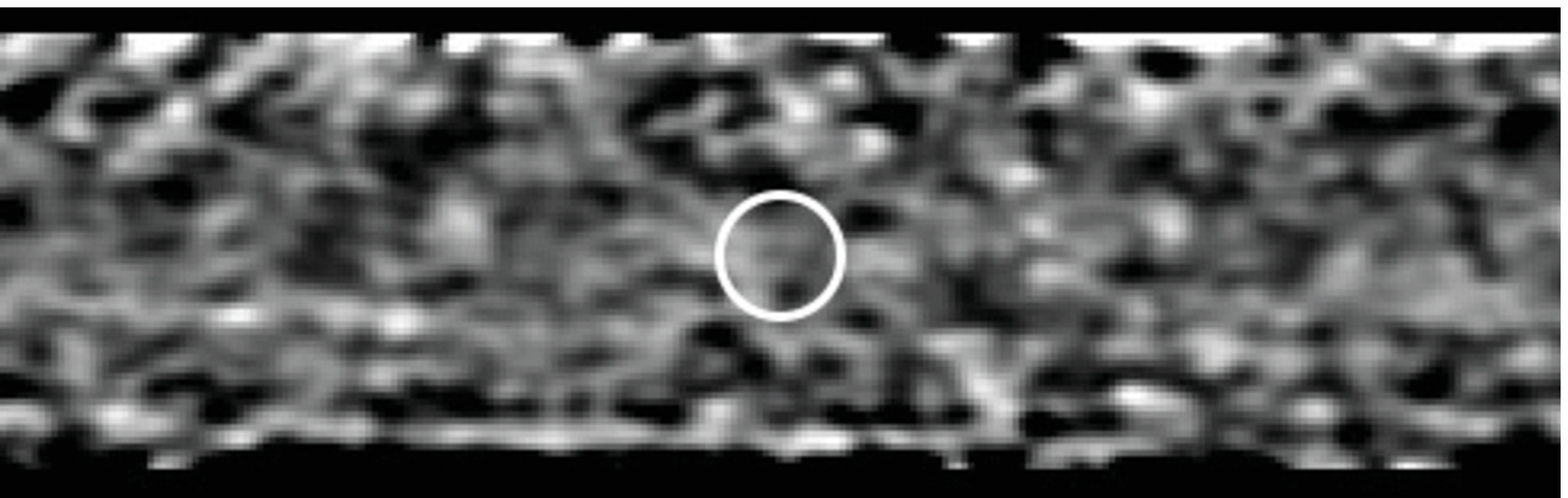}} \\
  \resizebox{0.47\textwidth}{!}{\includegraphics{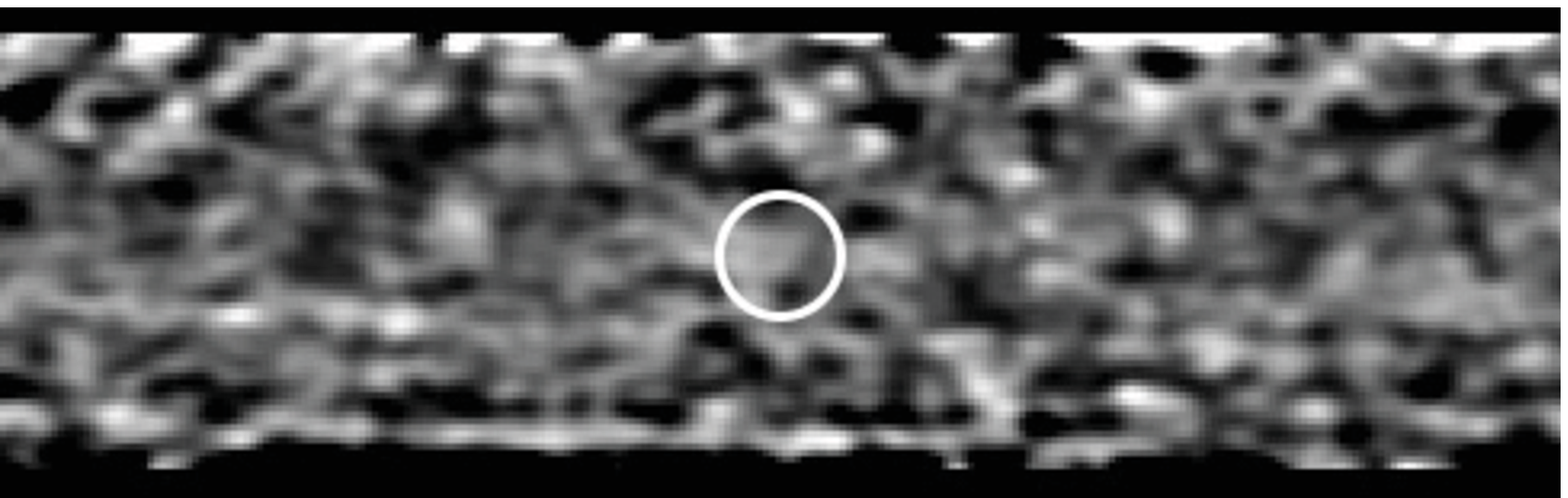}} \\
  \resizebox{0.47\textwidth}{!}{\includegraphics{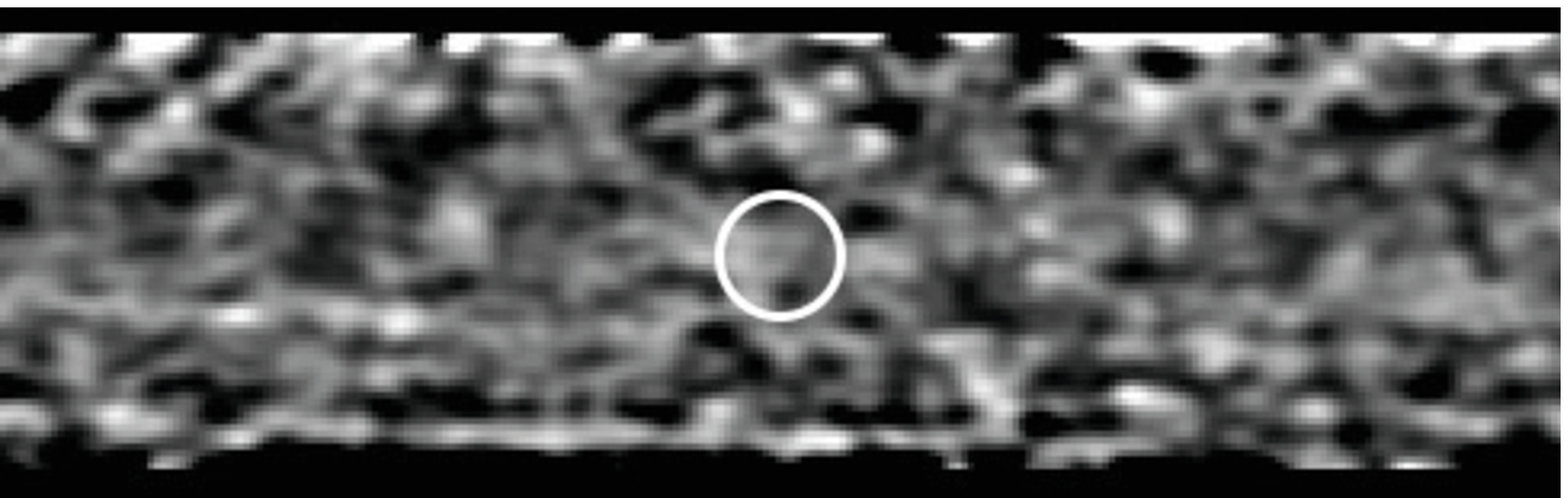}} \\
 \caption{The calibrated GNIRS spectrum, with the location
 of HUDF.zD1 and the expected wavelength of the tentative Lyman-$\alpha$ emission reported 
 by Fontana et al.\ (2010) marked with a white circle. Wavelength increases from left to right, and 
 we show the 95\,\AA\ either side of 9691.5\,\AA. 
 From top to bottom: (a) the reduced data. The three vertical lines of higher noise are due to night sky emission lines; (b) the reduced data 
 convolved with an elliptical Gaussian with $\sigma$
 of 1.4\,pixels spatially and 2.2\,pixels spectrally, matching the profile of a Gaussian emission line
 with FWHM of $0\farcs5$ and 2.6\,\AA .
 (c) a fake source with the same line flux ($3.4\times 10^{-18}\,{\rm erg\,cm^{-2}\,s^{-1}}$)
 and wavelength as the Fontana et al.\ (2010) line added into the frame. The resulting frame has been Gaussian smoothed. We assume
 a spatially and spectrally unresolved source. If the line is spectrally resolved, the S/N would
 be lower -- panel  (d) shows the expected 2D Gaussian-smoothed spectrum for
 an emission line with an intrinsic velocity width of 200\,km\,s$^{-1}$ (FWHM) and the same
 line flux as reported in Fontana et al.\ (2010). }
 \label{fig:fontana}
\end{figure}

\section{Analysis}
\label{sec:analysis}

\subsection{The EW distribution and Constraints on the Neutral Fraction $\chi_{\scriptscriptstyle HI}$}
\begin{figure}
   \resizebox{0.45\textwidth}{!}{\includegraphics{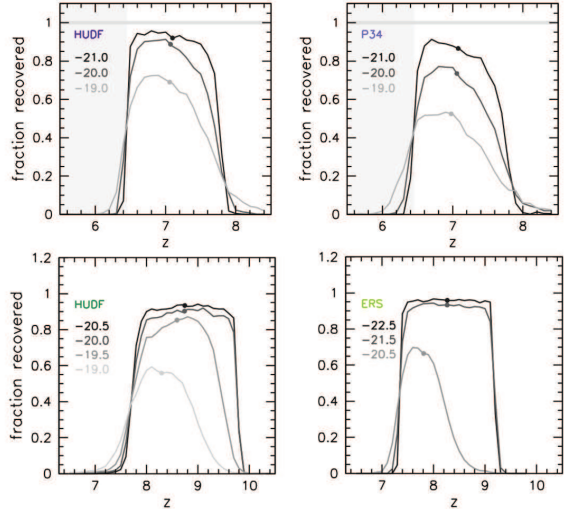}} \\
 \caption{The probability of recovering a galaxy in the simulations described in Wilkins et al.\ (2011) and Lorenzoni et al.\ (2011) for different fields as a function of redshift for several different absolute rest-frame M$_{1600}$ magnitudes.  In this paper, we use the results of these simulations to provide the expected redshift distributions. The upper two figures show the simulations for $z$-band dropouts whereas the lower two figures are for $Y$-band dropouts.  In each case, the mean redshift is denoted by a dot.} 
 \label{fig:sims}
\end{figure}

Given the fact that we detect no Lyman-$\alpha$ emission in any of our spectra, what can we say about the strength of Lyman-$\alpha$ emission in $z>7$ galaxies?  The broad-band filters used to colour-select candidates with the Lyman break technique introduce a selection function on the redshift.  The expected redshift distribution used for our sample is taken from the simulations described in Wilkins et al.\ (2011) at $z=7$ for the $z$-band dropouts and Lorenzoni et al.\ (2011) at $z=8$ for the $Y$-band dropouts, which are reproduced in Fig.\ref{fig:sims}.  We compute the probability of recovering a galaxy as a function of redshift for different rest-frame UV absolute magnitudes around $\lambda_{rest}=$\,1600\,\AA\ (M$_{1600}$).  For each spectroscopic target, we calculated M$_{1600}$ so that for every object we could determine the most appropriate scenario to use from the simulations in Fig.\ref{fig:sims}, choosing the relevant curve for a given M$_{1600}$ and field (eg. HUDF, P34 etc).  These curves give us the probability of recovering a galaxy at a given redshift.

We then considered different thresholds on the rest-frame EW (50\,\AA, 80\,\AA\, and 120\,\AA) for each of our targets.  For each particular target and chosen EW threshold, we computed the fraction of our spectroscopy that had EW limits lower than our chosen threshold (i.e. the fraction of the spectrum where EW$_{\mathrm{upper limit}} < $ EW$_{\mathrm{threshold}}$, which we call Frac$_{\mathrm{EW < thres}}$, weighted by the redshift probability distribution for the dropout galaxy).  

By considering the likelihood function for a galaxy to be lying at a particular redshift, we computed the fractional probability, Frac$_z$, that a galaxy drawn from the dropout sample would fall within the spectral coverage of that spectrograph setup (tabulated in column 3 of Table \ref{table:fractions}).  Multiplying Frac$_z$ by Frac$_{\mathrm{EW < thres}}$ (found in columns 5-7 of Table \ref{table:fractions}) and computing the sum over all the galaxies observed gives us the effective total number of galaxies, $N_{\mathrm{eff}}$, where our sensitivity is greater than our chosen EW threshold.  These are tabulated in Table \ref{table2}.

Given that we do not detect any Lyman-$\alpha$ emission in our data we now consider what scenarios of EW evolution from lower redshift we can rule out.  From Poisson statistics, if a given model predicts that on average $\lambda_{\rm ex}$ galaxies are expected to be detected in the survey, the probability $f_{n}$ of detecting $n$ galaxies is given by
\[
f_{n}=\frac{(\lambda_{\rm ex})^n e^{-\lambda_{\rm ex}}}{n!} .
\]
We detect no galaxies in our survey ($n=0$), and are able to reject models which predict $\lambda_{\rm ex}$ detections at the $(1-f_{n})$ level; i.e.\ a model which predicts 1 galaxy is rejected at the 63 per cent level (roughly corresponding to $1\,\sigma$ for a Normal distribution), and a model which predicts 3 galaxies are detected is rejected with 95\% confidence (corresponding to about $2\,\sigma$ for a Normal distribution).

\begin{figure}
    \resizebox{0.40\textwidth}{!}{\includegraphics{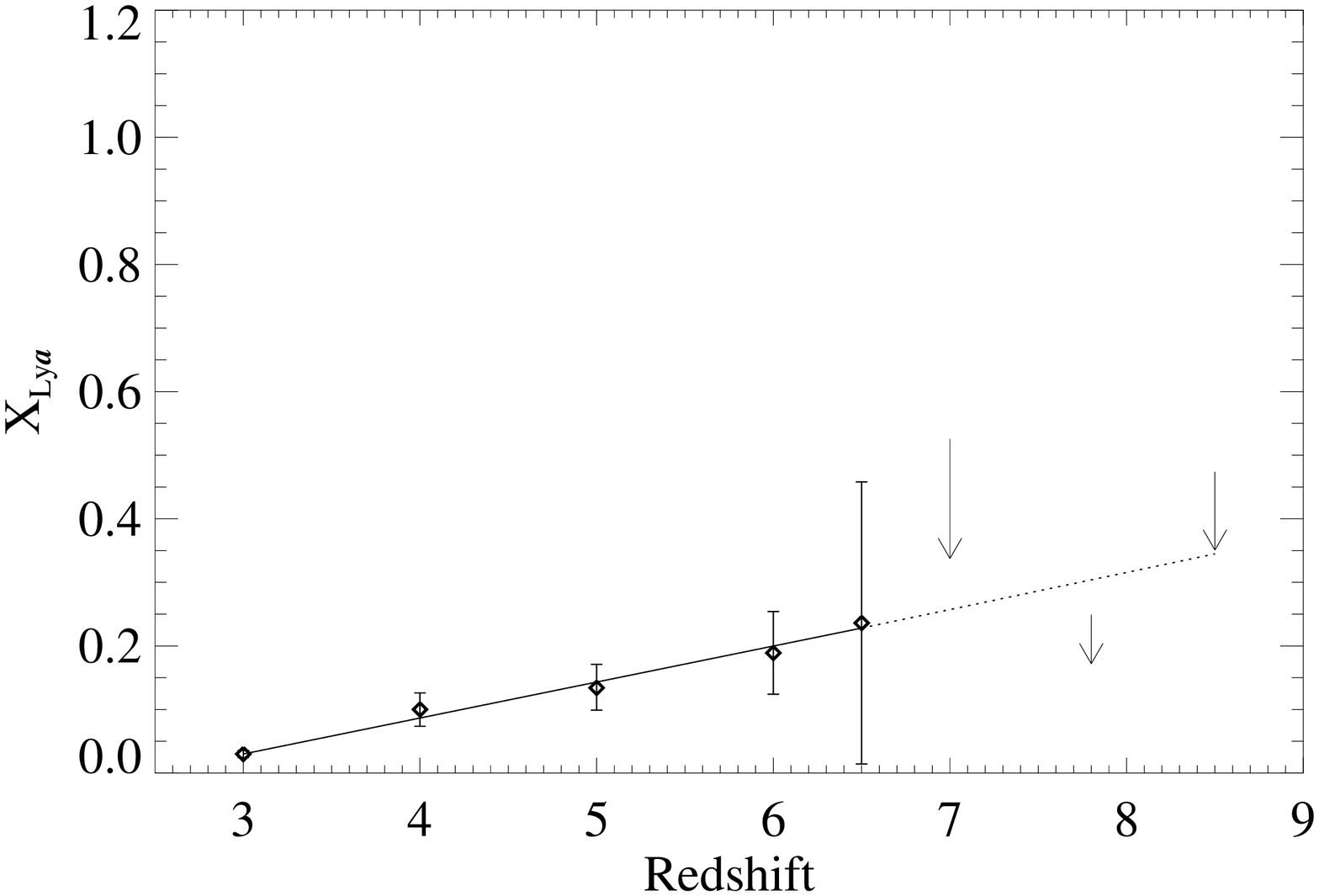}} \\
    \resizebox{0.40\textwidth}{!}{\includegraphics{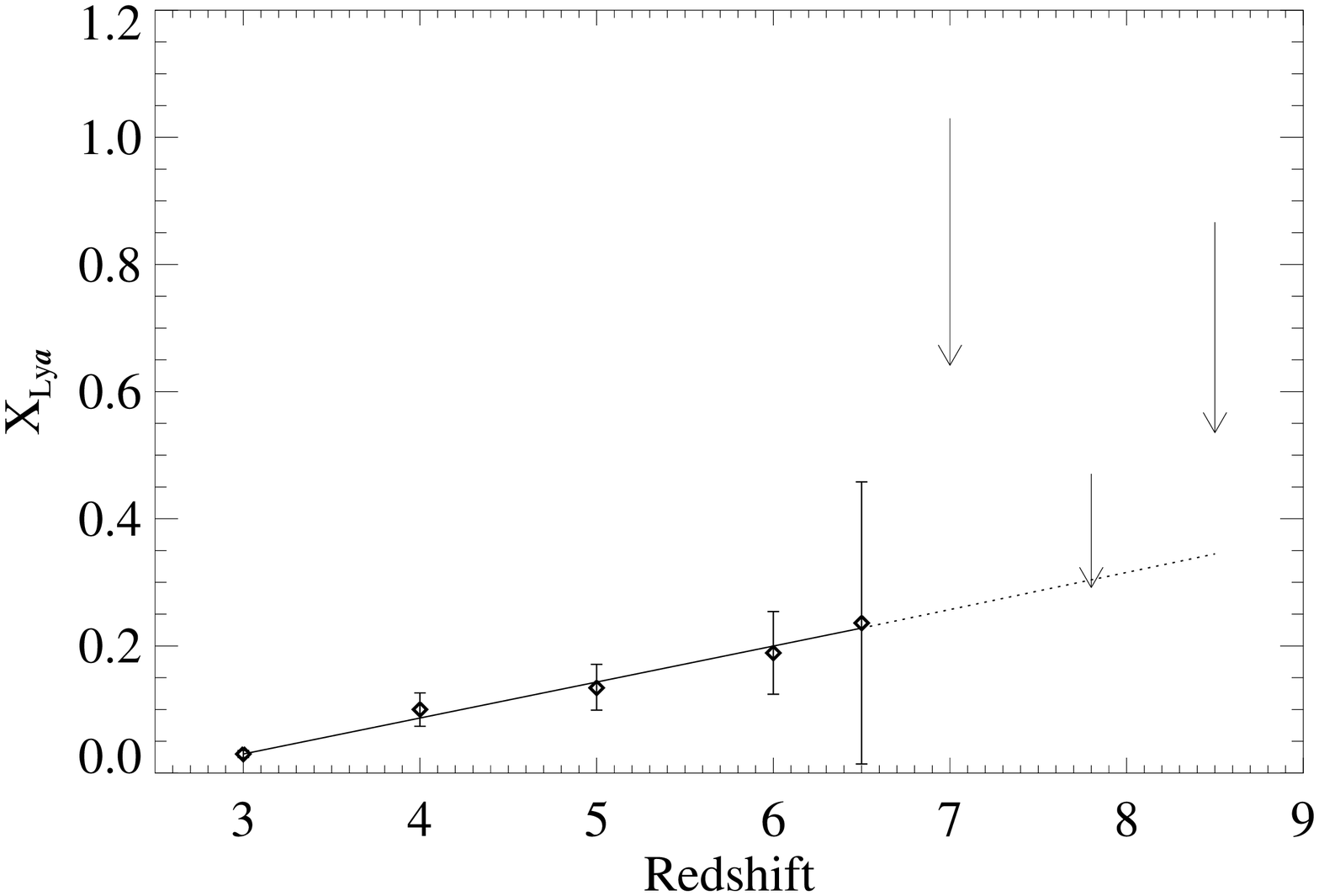}} \\
 \caption{Our upper limits on the fraction of high rest-frame equivalent Lyman-$\alpha$ emission at $z\ge7$ are shown for the $z$-drops ($z=7$), the $Y$-drops ($z=8.5$) and our complete sample (with mean $z=7.8$). The head of each arrow is the 120\,\AA\ limit and the tail of each arrow is the 75\,\AA\ limit. 
 The diamond symbols are results obtained at lower redshift by Shapley et al.\ (2003) at $z=3$ and Stark et al.\ (2010) at $z=4-6.5$. For comparison, we extrapolate the low-redshift trend to higher redshifts (dotted line). Our upper limits appear inconsistent with this extrapolation at the $1\,\sigma$ level, perhaps indicating that the IGM neutral fraction $\chi_{\scriptscriptstyle HI}>0$ at $z>7$.  The upper figure shows our constraints when considering an unresolved line, whereas the lower figure considers a line with an intrinsic velocity width of 200km/s.}
 \label{fig:constraint}
\end{figure}

Specifically, we can rule out at the 63 percent ($\approx 1\,\sigma$) level any scenario that predicts a fraction of galaxies with Lyman-$\alpha$ emission above the threshold EW, of $X_{Ly\alpha}>1/N_{\mathrm{eff}}$ (taking $\lambda_{\rm ex}=X_{Ly\alpha}\times N_{\mathrm{eff}}=1$  in the above equation).  Here we compare our results with the work of Stark et al.\ (2010), building on the previous work of Stanway et al.\ (2007) at $z=6$ and Shapley et al.\ (2003) at $z=3$. Stark et al.\ (2010) determined the fraction of dropout galaxies with EW $\gtrsim 75$\,\AA\ at lower redshift ($z=4 - 6.5$), and showed that over this redshift range the fraction of strong Lyman-$\alpha$ emitters increased with increasing redshift. The four data points with error bars in Figure \ref{fig:constraint} are from Stark et al.\ (2010) and show the fraction of strong Lyman-$\alpha$ emitters at different redshifts in their spectroscopic dropout sample, which covers a range in UV luminosities of $-19.5>M_{UV}>-20.5$, similar to the range of luminosities of our higher-redshift sample presented here.  We also add a point at $z=3$ from Shapley et al.\ (2003), who found that 29 of 957 $U$-band drop-out galaxies had $EW>80$\,\AA. We do a simple linear extrapolation in redshift of this trend out to $z=8.5$ (dotted line), where we are sensitive with the $Y$-drops.  
This scenario would correspond to constant evolution (linear with redshift) in the intrinsic Lyman-$\alpha$ EW distribution coupled with no evolution in the neutral fraction from lower redshifts, with $\chi_{\scriptscriptstyle HI}=0$.  The three upper-limit arrows in Figure \ref{fig:constraint} show the constraints derived from our observations; the tail and head of the arrows represent the expected fraction of Lyman-$\alpha$ emitting galaxies with EW$<$\,75\,\AA\ and EW$<$\,120\,\AA\ respectively.  The arrow at $z=7$ shows the constraint we get from our $z$-drops, whereas the one at $z=8.5$ shows the constraint from our $Y$-drops. The mean redshift for $z$-drops and $Y$-drops was derived from the simulations by Wilkins et al.\ (2011) and Lorenzoni et al.\ (2011). The arrow in the middle at $z=7.8$ is the constraint obtained by considering the $z$-drops and $Y$-drops together, at a mean redshift of $z=7.8$. For a spectrally-unresolved line, we would expect on average 1.2 galaxies to be detected in our combined sample if the extrapolated evolution of the Lyman-$\alpha$ fraction holds (i.e.\ $X_{Ly\alpha}\approx 0.3$ at $z=7.8$); we do not detect any galaxies, and hence the hypothesis is mildly inconsistent with our upper limits at the 70 per cent level. If the emission lines have intrinsic velocity widths of $\approx  200$\,km\,s$^{-1}$, then we would expect only 0.64 galaxies to have detectable Lyman-$\alpha$ emission in our sample adopting the extrapolated $X_{Ly\alpha}$, so we cannot rule out this scenario with our current data
 (it is formally inconsistent with our upper limit only at the 47 per cent level).

\subsection{Brief review of and comparison with other studies}

Efforts to obtain spectroscopic detection of Lyman-$\alpha$ emission at $z\sim7$ have also been made by other groups.  In this section, we summarize the results of these other studies and compare and contrast our own results.

Fontana et al.\ (2010) used VLT/FORS2 to observe 7 Lyman-break galaxy candidates selected in the GOODS-S field from Hawk-I/VLT and WFC3/{\em HST} imaging, and other than the one tentative Lyman-$\alpha$ emission line in G2\_1408 (HUDF.zD1 in the catalogue of Bunker et al.\ 2010) which we also target and has been discussed above, they detect no other Lyman-$\alpha$ emission in the rest of the sample.  

As part of the same survey, Vanzella et al.\ (2011) report the detection of Lyman-$\alpha$ emission in 2 objects in the BDF4 field (Lehnert \& Bremer 2003).  The results of the final sample were discussed in Pentericci et al.\ (2011), who observed 20 $z$-drop galaxies with VLT/FORS2 and detected 5 Lyman-$\alpha$ lines in their sample.  Adopting simulation techniques discussed in Fontana et al.\ (2010) they report that, on the basis of observations made at lower redshift, the probability of detecting only 5 galaxies in their sample is below 2\%, indicating a declining fraction of strong Lyman-$\alpha$ emitters at higher redshifts, consistent with our results.

Schenker et al.\ (2012) presented Keck LRIS observations of 19 sources with photometric redshifts lying in the range $6.3 < z < 8.8$.  They reported two convincing Lyman-$\alpha$ detections (ERS 8496 at $z=6.441$ and A1703\_zD6 at $z=7.045$), and a marginal detection at $z=6.905$ for HUDF09\_1596. 
For their discussion, they added the 7 objects discussed by Fontana et al.\ (2010) such that they could carry out an analysis over a larger sample of objects. They conclude that 24 of their targets (out of 26) have spectral coverage over more than half the likely wavelength range for Lyman-$\alpha$ (given the photometric redshifts using the EAZY code).  From simulations they show that out of the combined 26 targets observed by them and Fontana et al.\ (2010), they should have detected 7-8 emission lines rather than just 2, again arguing for a steep decline in the Lyman-$\alpha$ fraction at $z>6.3$.

Ono et al.\ (2012) carried out observations of 11 $z$-drops in the SDF and GOODS-N using Keck/DEIMOS, detecting Lyman-$\alpha$ emission for 3 objects in their sample, one of which had already been detected at a lower level of significance by Iye et al.\ (2006).

All of the objects discussed above were brighter in the $Y$-band than any of our spectroscopic targets.  We checked whether we would have detected Lyman-$\alpha$ emission from any of these objects in our own spectroscopy if we had targeted them.  To perform these comparisons we first considered an appropriate line profile for each individual object; in each case, the emission was assumed to be spatially unresolved, whereas for the spectral extent of the lines we adopted the published values and convolved these with the spectral resolution of our instruments\footnote{A1703\_zD6 was unresolved in the observations of Schenker et al.\ using NIRSPEC ($\Delta\lambda_{\mathrm{FWHM}}=6.5$\AA), so here we are assuming that this object would be unresolved in our observations with GNIRS and XSHOOTER.}.  We chose our aperture to be about twice the size of the adopted emission line profile (spatially and spectrally) and performed aperture corrections to allow for any loss in flux in our apertures.  We report the results of these comparisons in Table \ref{table:comparisons}.  In summary, we conclude that nearly all objects with the line fluxes detected by other groups would have been clearly detected in our spectroscopy\footnote{As can be seen from Table \ref{table:comparisons}, the only two objects which we would not have detected in either GNIRS or XSHOOTER are NTTDF-6345 and NTTDF-474, whose Lyman-$\alpha$ flux is measured to be quite low.  ERS 8496 (which also has a relatively low measured Lyman-$\alpha$ flux) would have been undetected in our GNIRS observations, but would have been clearly detected in our XSHOOTER data.  BDF-3299, on the other hand, would have been detected in our GNIRS observations, but not in our XSHOOTER data.}.  However, we note that the detections presented by other groups are typically for objects with brighter broad-band magnitudes, and hence the equivalent widths are small. Only 4 of the 9 objects in Table~\ref{table:comparisons} have rest-frame equivalent widths $\ge50$\,\AA , where our observations are sensitive over an appreciable redshift range. The others have smaller rest-frame equivalent widths, which would be probed only by our spectra of P34.z.4809 and zD2 (around $1.05\mu$m). This suggests that whilst there might indeed be no Lyman-$\alpha$ emission escaping the objects which we targetted in our spectroscopy, the negative results could also be due the the possibility of the emission being fainter than our detection limits, particularly given that our targetted objects are fainter than any of the other objects in literature discussed above. It has been suggested, however, that the fraction of high equivalent-width Lyman-$\alpha$ emitters is larger for faint galaxies (Stark et al.\ 2010, 2011), which means that our 
 derived upper limits on the equivalent width of these objects, shown in figures \ref{fig:EWlimitzD1} - \ref{fig:P34Z4809}, may offer stronger constraints on $\chi_{HI}$.

\begin{table*}
\caption{This table shows the significance levels at which objects discussed in literature would have been detected in our spectroscopy if we had targetted them with Gemini/GNIRS or VLT/XSHOOTER.}
\begin{tabular}{| c | c | c | c | c | c | c |}
\hline
Object & Redshift & Mag (Y$_{AB}$) & Lyman-$\alpha$ flux & EW / \AA & GNIRS $\sigma$ & XSHOOTER $\sigma$ \\ 
& & & (erg cm$^{-2}$s$^{-1}$) & & \\
\hline
\hline
ERS 8496 & $6.441 \pm 0.002$ & N/A & $0.91\pm1.4\times10^{-17}$ & $69\pm 10$ & 1.52 & 6.5 \\
\hline
A1703\_zD6 & 7.045 & N/A & $2.84 \pm 5.3 \times 10^{-17}$ & $65\pm12$ & 18.93 & 16.7 \\
\hline
BDF-521 & $7.008 \pm 0.002$ & 25.86 & $1.6\pm0.16\times10^{-17}$ & $64^{+10}_{-9}$ & 6.67 & 5.93 \\
\hline
BDF-3299 & $7.109 \pm 0.002$ & 26.15 & $1.2\pm0.14\times10^{-17}$ & $50^{+11}_{-8}$ & 4.80 & 2.00 \\
\hline
GN-108036 & 7.213 & 25.50 & $2.5\times10^{-17}$ & 33 & 5.00 & 3.57 \\ 
\hline
SDF-63544 & 6.965 & 25.10 & $2.8\times10^{-17}$ & 43 & 8.62 & 5.19 \\ 
\hline
SDF-46975 & 6.844 & 25.20 & $2.7\times10^{-17}$ & 43 & 7.20 & 13.50 \\
\hline
NTTDF-6345 & 6.701 & 25.46 & $0.72\times10^{-17}$ & $15\pm3$ & 0.48 & 2.4 \\
\hline
NTTDF-474 & 6.623 & 26.50 & $0.32\times10^{-17}$ & $16\pm5$ & 0.64 & 1.78 \\
\hline
\end{tabular} 
\label{table:comparisons} 
\end{table*}

\section{CONCLUSIONS}
We have presented spectroscopic observations with GEMINI/GNIRS and VLT/XSHOOTER of a sample of $z > 7$ candidate galaxies and we fail to detect significant Lyman-$\alpha$ emission from any of them.  This is consistent with the fraction of high rest-frame equivalent width Lyman-$\alpha$ emitters dropping at $z>7$, as would be expected if the neutral HI fraction was greater at these epochs.  We have also investigated a tentative emission line published by Fontana et al.\ (2010) in HUDF.zD1 (from the catalogue of Bunker et al.\ 2010) and our analysis does not confirm the presence of this line, although we do not rule out the possibility of it being real, especially if it has considerable velocity extent.

Given the lack of Lyman-$\alpha$ emission in our spectroscopy in conjunction with the continuum flux derived from {\em HST} imaging of these objects, we derived upper limits on the rest-frame equivalent width of our objects.  Extrapolating the Lyman-$\alpha$ fraction observed at lower redshifts by Stark et al.\ (2010) and Shapley et al.\ (2003), our lack of Lyman-$\alpha$ detection rules out at a level of 1\,$\sigma$ (70\%), for spectrally unresolved lines, the scenario in which the Lyman-$\alpha$ fraction evolves with the same trend found at lower redshifts.  The limits are weaker if the lines have significant velocity width extent.  A diminished Lyman-$\alpha$ fraction at higher redshift is consistent with other published studies.  This attenuation in the Lyman-$\alpha$ fraction can be attributed either to physical evolution of the galaxies or, more likely, an increase in the neutral fraction of hydrogen at $z > 7$, i.e. these observations can most likely be interpreted as implying that the neutral fraction at $z\sim8$ can be ruled out as being $\chi_{\scriptscriptstyle HI}=0$ at a level of 1\,$\sigma$.  Larger-number statistics are required to confirm this hypothesis at a higher level of significance.  To this end, we have undertaken spectroscopy on a large sample of $z$-band and $Y$-band dropouts with VLT/FORS2 and SUBARU/MOIRCS (Caruana et al.\ {\em in prep}).

\label{sec:conc}

\subsection*{Acknowledgements}
Based on observations obtained at the Gemini Observatory, which is operated by the Association of Universities for
    Research in Astronomy, Inc., under a cooperative agreement with the
    NSF on behalf of the Gemini partnership: the National Science
    Foundation (United States), the Science and Technology Facilities
    Council (United Kingdom), the National Research Council (Canada),
    CONICYT (Chile), the Australian Research Council (Australia),
    Minist\'{e}rio da Ci\^{e}ncia, Tecnologia e Inova\c{c}\~{a}o (Brazil)
    and Ministerio de Ciencia, Tecnolog\'{i}a e Innovaci\'{o}n Productiva
    (Argentina).  The Gemini programmes associated with these observations were:  GS-2004B-Q-19 and GS-2005B-Q-18.

Based on observations made with the NASA/ESA {\em Hubble} Space Telescope associated with programmes \#GO-11563, \#GO/DP-10086, \#GO-9803, \#GO-9425.
obtained from the Data Archive at the Space Telescope Science Institute, which is operated by the Association
of Universities for Research in Astronomy, Inc., under NASA contract
NAS 5-26555. 

We gratefully acknowledge Chris Willott, Richard Ellis, Daniel Stark and the NIRSpec Instrument Science Team for useful discussions.  We thank the anonymous referee whose helpful and insightful comments greatly improved this manuscript.

MJJ acknowledges the support of a RCUK
fellowship. 
JC and SL are supported by the Marie Curie Initial Training Network ELIXIR of the European Commission under
contract PITN-GA-2008-214227.
AB and SW acknowledge financial support from an STFC Standard Grant.

\begin{table*}
\caption{This table for all objects targeted by our spectroscopy shows the redshift range spanned by our data for Lyman-$\alpha$ in Column 2.  Column 3 shows the fractional probability that a galaxy drawn from the dropout sample would fall within the spectral coverage of that particular spectrograph setup.  Column 4 gives the median EW for each object for the most probable redshift range (see Figure \ref{fig:sims}).  The remaining three columns tabulate the fraction of our spectroscopy that has EW limits lower than our chosen threshold (50\,\AA, 75\,\AA\, and 120\,\AA\, respectively.)  The figures in this table are for an unresolved line.}
\begin{tabular}{ | c | c | c | c | c | c | c | }
\hline
Object & z-Range spanned & Frac$_{z}$ & Median EW & Frac$_{\mathrm{EW < 50\AA}}$ &  Frac$_{\mathrm{EW < 75\AA}}$ & Frac$_{\mathrm{EW < 120\AA}}$\\ 
	    & by data (for Ly-$\alpha$)& & & & & \\
\hline
\hline
zD1 			& 5.82 -- 6.40 & 0.01 & 54.69 & 0 & 0.4583 & 1.0000 \\
			& 6.08 -- 6.74 & 0.24 & 243.2 & 0 & 0 & 0 \\
			& 6.79 -- 7.45 & 0.50 & 29.35 & 0.8349 & 0.9415 & 0.9990 \\
			& 7.25 -- 8.05 & 0.36 & 85.29 & 0 & 0.3326 & 0.6299 \\
\hline
zD2			& 6.09 -- 6.73 & 0.24 & 113.5 & 0 & 0 & 0.7697 \\
 			& 7.25 -- 8.00 & 0.33 & 76.10 & 0.285052 & 0.4109 & 0.6883 \\
\hline
zD3			& 6.09 -- 6.73 & 0.24 & 120.4 & 0 & 0 & 0.7262 \\
			& 7.25 -- 8.00 & 0.33 & 80.72 & 0 & 0.3601 & 0.6413 \\
\hline
zD4 			& 6.09 -- 6.73 & 0.24 & 150.2 & 0 & 0 & 0.1869 \\
			& 7.25 -- 8.00 & 0.33 & 100.8 & 0 & 0.0689 & 0.5326 \\
\hline
P34.z.4809	& 3.60 -- 7.40 (Optical) & 0.70 & 19.45 & 0.8268 & 0.9188 & 0.9856  \\
& 7.42 - 19.40 (NIR) & 0.30 & 35.21 & 0.8191 & 1.2917 & 1.7287  \\
\hline
ERS.YD2		& 7.42 -- 19.40 (NIR)  & 0.86 & 67.19 & 0.6529 & 1.1551 & 1.4293 \\
\hline
HUDF.YD3	& 7.42 -- 19.40  (NIR) & 0.98 & 241.6 & 0.0550 & 0.1881 & 0.6516 \\
\hline
UDF092y-03751196d & 7.42 -- 19.40 (NIR) & 0.996 &  28.64 & 0.8400 & 0.9374 & 0.9872 \\	
\end{tabular} 
\label{table:fractions} 
\end{table*}

\begin{table*}
\caption{This table shows the total effective number of sampled galaxies with an EW upper limit lower than a set threshold.  We present these figures separately for $z$-drops, $Y$-drops and both $z$-drops and $Y$-drops combined.  The figures in this table are for an unresolved line.}
\begin{tabular}{| c | c | c | c | c |}
\hline
& Average Redshift & & $N_{\mathrm{eff}}=\sum$ Frac$_{z} \times$ Frac$_{\mathrm{EW < thres}}$  & \\
& & & & \\
& & EW$_{\mathrm{thres}}=50$\,\AA & EW$_{\mathrm{thres}}=75$\,\AA & EW$_{\mathrm{thres}}=120$\,\AA \\
\hline
z-drops & 7.0 & 1.242 & 1.903 & 2.963 \\
\hline
$Y$-drops & 8.5 & 1.452 & 2.111 & 2.851 \\
\hline
$z$-drops \& $Y$-drops & 7.8 & 2.694 &  4.014 & 5.814 \\
\hline
\end{tabular}
\label{table2}
\end{table*}


\begin{table*}
\caption{This table for all objects targeted by our spectroscopy shows the redshift range spanned by our data for Lyman-$\alpha$ in Column 2.  Column 3 shows the fractional probability that a galaxy drawn from the dropout sample would fall within the spectral coverage of that particular spectrograph setup.  Column 4 gives the median EW for each object for the most probable redshift range (see Figure \ref{fig:sims}).  The remaining three columns tabulate the fraction of our spectroscopy that has EW limits lower than our chosen threshold (50\,\AA, 75\,\AA\, and 120\,\AA\, respectively.) The figures in this table are for a 200km/s line.}
\begin{tabular}{ | c | c | c | c | c | c | c | }
\hline
Object & z-Range spanned & Frac$_{z}$ & Median EW & Frac$_{\mathrm{EW < 50\AA}}$ &  Frac$_{\mathrm{EW < 75\AA}}$ & Frac$_{\mathrm{EW < 120\AA}}$\\ 
	    & by data (for Ly-$\alpha$)& & & & & \\
\hline
\hline
zD1 			& 5.82 -- 6.40 & 0.01 & 107.4 & 0 & 0 & 0 \\
			& 6.08 -- 6.74 & 0.24 & 476.2 & 0 & 0 & 0 \\
			& 6.79 -- 7.45 & 0.50 & 69.38 & 0.1958 & 0.5460 & 0.8672 \\
			& 7.25 -- 8.05 & 0.36 & 167.3 & 0 & 0 & 0.1016 \\
\hline
zD2			& 6.09 -- 6.73 & 0.24 & 222.5 & 0 & 0 & 0 \\
 			& 7.25 -- 8.00 & 0.33 & 149.3 & 0 & 0 & 0.2177 \\
\hline
zD3			& 6.09 -- 6.73 & 0.24 & 235.9 & 0 & 0 & 0 \\
			& 7.25 -- 8.00 & 0.33 & 158.4 & 0 & 01 & 0.1062 \\
\hline
zD4 			& 6.09 -- 6.73 & 0.24 & 294.4 & 0 & 0 & 0 \\
			& 7.25 -- 8.00 & 0.33 & 197.7 & 0 & 0 & 0 \\
\hline
P34.z.4809	& 3.60 -- 7.40 (Optical) & 0.70 & 38.87 & 0.5759 & 0.6811 & 0.8617  \\
& 7.42 -- 19.40 (NIR) & 0.30 & 58.65 & 0.2650 & 0.7361 & 1.2627  \\
\hline
ERS.YD2		& 7.42 -- 19.40 (NIR)  & 0.86 & 112.4 & 0.1583 & 0.4344 & 1.0600 \\
\hline
HUDF.YD3	& 7.42 -- 19.40  (NIR) & 0.98 & 400.3 & 0 & 0 & 0.0431 \\
\hline
UDF092y-03751196d & 7.42 -- 19.40  (NIR) & 0.996 &  50.63 & 0.5331 & 0.7838 & 0.9181 \\	
\end{tabular} 
\label{table:fractions_200kms} 
\end{table*}

\begin{table*}
\caption{This table shows the total effective number of sampled galaxies with an EW upper limit lower than a set threshold.  We present these figures separately for $z$-drops, $Y$-drops and both $z$-drops and $Y$-drops combined. The figures in this table are for a 200km/s line.}
\begin{tabular}{| c | c | c | c | c |}
\hline
& Average Redshift & & $N_{\mathrm{eff}}=\sum$ Frac$_{z} \times$ Frac$_{\mathrm{EW < thres}}$  & \\
& & & & \\
& & EW$_{\mathrm{thres}}=50$\,\AA & EW$_{\mathrm{thres}}=75$\,\AA & EW$_{\mathrm{thres}}=120$\,\AA \\
\hline
z-drops & 7.0 & 0.581 & 0.971 & 1.559 \\
\hline
$Y$-drops & 8.5 & 0.667 & 1.154 & 1.868 \\
\hline
$z$-drops \& $Y$-drops & 7.8 & 1.248 & 2.125 & 3.427 \\
\hline
\end{tabular}
\label{table2_200kms}
\end{table*}

\bsp


\begin{thebibliography}{}
\bibitem[\protect\citename{Becker et al.\ } 2001]{be01}
Becker R.~H. et al., 2001, AJ, 122, 2850

\bibitem[\protect\citename{Beckwith et al.\ } 2006]{be06}
Beckwith S.~V.~W et al., 2006, AJ, 132, 1729

\bibitem[\protect\citename{Bertin \& Arnouts} 1996]{be96}
Bertin E., Arnouts S., 1996, A\&AS, 117, 393

\bibitem[\protect\citename{Bouwens et al.\ } 2006]{bo06}
Bouwens R.~J. Illingworth G.~D. Blakeslee J.~P.; Franx M., 2006, ApJ, 653, 53

\bibitem[\protect\citename{Bouwens et al.\ } 2007]{bo07}
Bouwens R.~J., Illingworth G.~D., Franx M. \& Ford H., 2007, ApJ, 670, 928

\bibitem[Bouwens et al.(2010)]{2010ApJ...709L.133B} Bouwens, R.~J., et al.\ 
2010, \apjl, 709, L133 

\bibitem[Bouwens et al.(2011)]{2011ApJ...737:90} Bouwens, R.~J., et al.\
2011, \apjl, 737:90

\bibitem[Bunker et al.(2003)]{2003MNRAS.342L..47B}
Bunker, A. ~J., Stanway, E. R., Ellis, R. S., McMahon, R. G. \& McCarthy, P. J. 2003, \mnras, 342, L47

\bibitem[\protect\citename{Bunker et al.\ } 2004]{bu04}
Bunker A.~J., Stanway E.~R., Ellis R.~S., McMahon R.~G., 2004, MNRAS, 355, 374

\bibitem[Bunker et al.(2010)]{2010MNRAS.409..855B} Bunker, A.~J., et al.\ 
2010, \mnras, 409, 855 

\bibitem[D'Odorico et al.(2006)]{2006SPIE.6269E..98D} D'Odorico, S., et 
al.\ 2006, Proc.\ SPIE, 6269,  

\bibitem[Finkelstein et al.(2010)]{Finkelstein}
Finkelstein, S. L., Papovich, C., Giavalisco, M., Reddy, N. A.,
Ferguson, H. C., Koekemoer, A. M., \& Dickinson, M. 2010,
ApJ, 719, 1250

\bibitem[Fontana et al.(2010)]{2010ApJ...725L.205F} Fontana, A., et al.\ 
2010, \apjl, 725, L205 

\bibitem[Gunn 
\& Peterson(1965)]{1965ApJ...142.1633G} Gunn, J.~E., \& Peterson, B.~A.\ 1965, ApJ, 142, 1633 

\bibitem[Hayes et al.(2012)]{Hayes2012}
Hayes, M., Laporte, N., Pell{\'o}, R., Schaerer, D. \& Le Borgne, J-F, 2012arXiv1205.6815

\bibitem[Iye et al.(2006)]{2006Natur.443..186I} Iye, M., et al.\ 2006, 
\nat, 443, 186 

\bibitem[Lehnert et al.(2003)]{2003ApJ...593..630L}
Lehnert, M.~D. \& Bremer, M. 2003, ApJ, 593, 630

\bibitem[Lehnert et al.(2010)]{2010Natur.467..940L} Lehnert, M.~D., et al.\ 
2010, \nat, 467, 940 

\bibitem[Lorenzoni et al.(2011)]{2011MNRAS.414.1455L} Lorenzoni, S., 
Bunker, A.~J., Wilkins, S.~M., Stanway, E.~R., Jarvis, M.~J., 
\& Caruana, J.\ 2011, \mnras, 414, 1455 


\bibitem[McLure et al.(2010)]{2010MNRAS.403..960M} McLure, R.~J., Dunlop, 
J.~S., Cirasuolo, M., Koekemoer, A.~M., Sabbi, E., Stark, D.~P., Targett, 
T.~A., \& Ellis, R.~S.\ 2010, \mnras, 403, 960 

\bibitem[McLure et al.(2011)]{2011MNRAS.418..2074M}
McLure, R. J., et al.\ 2011, \mnras, 418, 2074

\bibitem[Modigliani et al.(2010)]{2010SPIE.7737E..56M} Modigliani, A., et 
al.\ 2010, Proc.\ SPIE, 7737

\bibitem[\protect\citename{Oke \& Gunn} 1983]{ok83}
Oke J.~B., Gunn J.~E., 1983, ApJ, 266, 713

\bibitem[Oesch et al.(2010a)]{2010ApJ...709L..16O}
Oesch, P. A. et al.\ 2010a, ApJ, 709, L16

\bibitem[Oesch et al.(2010b)]{2010ApJ...709L..21O}
Oesch, P. A. et al.\ 2010b, ApJ, 709, L21


\bibitem[Ono et al.(2012)]{2012ApJ...744...83O}
Ono, Y. et al.\ 2012, ApJ, 744, 83

\bibitem[Pentericci et al.(2011)]{2011ApJ...743..132P}
Pentericci, L. et al.\ 2011, ApJ, 743, 132

\bibitem[Schenker et al.(2012)]{2012ApJ...744..179S}
Schenker, M.~A. et al.\ 2012, ApJ, 744, 179

\bibitem[Shapley et al.(2003)]{2003ApJ...588...65S}
Shapley, A.~E., Steidel, C.~C. Pettini, M. \& Adelberger, K.~L. 2003 ApJ, 	588, 65

\bibitem[Stanway et al.(2004)]{2004ApJ...604L..13S} Stanway, E.~R., 
Glazebrook, K., Bunker, A.~J., et al.\ 2004a, \apjl, 604, L13 

\bibitem[Stanway et al.(2004)]{2004ApJ...607..704S} Stanway, E.~R., Bunker, 
A.~J., McMahon, R.~G., et al.\ 2004b, \apj, 607, 704 

\bibitem[Stanway et al.(2007)]{2007MNRAS.376..727S}
Stanway, E.~R. et al.\ 2007, \mnras, 376, 727

\bibitem[Stark et al.(2010)]{2010MNRAS.408.1628S} Stark, D.~P., Ellis, 
R.~S., Chiu, K., Ouchi, M., \& Bunker, A.\ 2010, \mnras, 408, 1628 

\bibitem[Stark et al.(2011)]{2011ApJ...728L...2S}
Stark, D.~P., Ellis, R.~S. \& Ouchi, M. 2011, ApJ, 728, L2

\bibitem[Vanzella et al.(2011)]{2011ApJ...730L..35V} Vanzella, E., et al.\ 
2011, \apjl, 730, L35 

\bibitem[van Dokkum(2001)]{2001PASP..113.1420V} van Dokkum, P.~G.\ 2001, 
\pasp, 113, 1420 

\bibitem[Wilkins et al.(2010)]{2010MNRAS.403..938W}
Wilkins, S.~M. et al.\ 2010, \mnras, 403, 938

\bibitem[Wilkins et al.(2011)]{2011MNRAS.411...23W}
Wilkins, S.~M., Bunker, A.~J., Lorenzoni, S., \& Caruana, J., \ 2011, \mnras, 411, 23W

\end{thebibliography}
\end{document}